\documentclass[extra,article,onecolumn]{./gji-latex/gji}
\usepackage[papersize={8.5in,11in},margin=1in]{geometry} 
 \expandafter\let\csname equation*\endcsname\relax
 \expandafter\let\csname endequation*\endcsname\relax
\usepackage{bm}
\usepackage{enumerate}
\usepackage{mathrsfs,color}
\usepackage{amsfonts}
\usepackage{amssymb}
\usepackage{amsmath}
\usepackage{float}
\usepackage{graphicx,afterpage}
\usepackage{epstopdf}
\usepackage[pdftex,colorlinks=true]{hyperref}
\usepackage{upgreek}
\usepackage[x11names]{xcolor} 
\usepackage{bm}

\usepackage{multirow}

\usepackage{tikz}

\hypersetup{
     colorlinks   = true,
     citecolor     = blue, % used blue to ease my editing
     linkcolor     = blue
}

\usepackage{lineno}
\newcommand{\EqRef}[1]{(\ref{#1})}
\newcommand{\ACC}{\text{ACC}}
\newcommand{\TP}{\text{TP}}
\newcommand{\FP}{\text{FP}}
\newcommand{\TN}{\text{TN}}
\newcommand{\FN}{\text{FN}}
\newcommand{\Neg}{\text{N}}
\newcommand{\Pos}{\text{P}}
\newcommand{\MCC}{\text{MCC}}
\newcommand{\F}{\text{F}_1}
\newcommand{\TPR}{\text{TPR}}
\newcommand{\FPR}{\text{FPR}}

\newcommand{\lmax}{\ell_{\mbox{\footnotesize max}}}
\newcommand{\mmax}{m_{\mbox{\footnotesize max}}}

\begin{document}
\title[Predicting dipole reversals]{Can one use Earth's magnetic axial dipole field intensity to predict reversals?}
\author[K.~Gwirtz, M.~Morzfeld, A.~Fournier, G.~Hulot]{K.~Gwirtz$^*$, M.~Morzfeld$^*$, A.~Fournier$^o$, G.~Hulot$^o$\\
$^*$ Institute of Geophysics and Planetary Physics, 
Scripps Institution of Oceanography,
University of California, San Diego;\\
$^o$ Universit\'e de Paris, Institut de Physique du Globe de Paris,
CNRS, F-75005 Paris, France.}

\maketitle
\begin{abstract}
\noindent
We study predictions of reversals of Earth's axial magnetic dipole field
that are based solely on the dipole's intensity.
The prediction strategy is, roughly, that
once the dipole intensity drops below a threshold, 
then the field will continue to decrease and a reversal (or a major excursion) will occur.
We first present a rigorous definition of an intensity threshold-based prediction strategy
and then describe a mathematical and numerical framework to investigate its validity and robustness
in view of the data being limited.
We apply threshold-based predictions to a hierarchy of numerical models, 
ranging from simple scalar models to 3D geodynamos.
We find that the skill of threshold-based predictions varies across the model hierarchy.
The differences in skill can be explained by differences in how reversals occur:
if the field decreases towards a reversal slowly (in a sense made precise in this paper),
the skill is high, and if the field decreases quickly, the skill is low. 
Such a property could be used as an additional criterion to identify which models qualify as Earth-like. 
Applying threshold-based predictions to 
Virtual Axial Dipole Moment (VADM) paleomagnetic reconstructions (PADM2M and Sint-2000)
covering the last two million years,
reveals a moderate skill of  threshold-based predictions for Earth's dynamo.
Besides all of their limitations, threshold-based predictions
suggests that no reversal is to be expected within the next 10 kyr.
Most importantly, however, we show that considering an intensity threshold 
for identifying upcoming reversals is intrinsically limited by the dynamic behavior of Earth's magnetic field. 
\end{abstract}

\section{Introduction}
Earth possesses a time-varying magnetic field which is generated and sustained 
by turbulent flow of liquid metal alloy in the core.
The field varies over a wide range of spatial and temporal scales,
but this paper focuses on the dynamics of the axial dipole component  
over millions of years (Myr henceforth), 
which are relevant for the investigation of dipole reversals.
When a reversal occurs, the intensity of the dipole collapses
and then builds up in reversed polarity, with the magnetic north pole
becoming the south pole and vice versa.
%When an excursion occurs, 
%the intensity of the dipole collapses to a small value,
%but then builds up with the same polarity as before.
%Focusing for the moment on dipole reversals,
%We note that the
Occurrence of dipole reversals is well-documented over the past 150 Myr 
\citep{O12, CK95,LK04}.
We thus know that the last reversal occurred about 780 kilo years (kyr henceforth) ago
and that the average reversal rate over the past 5-10 Myr is about 4 reversals per Myr
\citep[see, e.g.,][]{MB19}.
Given these numbers, 
we wonder if we can reliably predict
if a reversal can be expected to occur any time soon.
%when the next dipole reversal will occur.

At first sight, the task seems hopeless 
because simulations of Earth's magnetic field suggest
that the geomagnetic field is not predictable beyond a century
\citep{Hulot:2010, Lhuillier:2011}.
The typical time elapsed between two reversals is much larger (often hundreds of millennia)
which implies that the exact timing of a reversal cannot be predicted until a 
reversal is just about to happen \citep[see also][]{Hulot:1994,lhuillier2011grl}. 
This predictability limit, however, concerns the field in its full detail,
and one may be able to identify macroscopic conditions that occur over long timescales
that are largely independent of the detailed morphology of the field.
For the remainder of this paper, we assume that the predictability limit for reversals, 
viewed as macroscopic features,
is larger than the predictability limit of the field's details.
This is motivated by the rich low-frequency dynamics
of the long-term dipole field \citep{CJ05},
and by the recent study of \citet{MFH17} that suggested this could possibly be the case.

In fact, many researchers have implicitly relied on this assumption
and studied precursors of dipole reversals.
Examples include careful investigations of the characteristics of past reversals
\citep[see, e.g.,][for a recent review]{valet2016deciphering},
studying the field structure during reversals and excursions \citep{BKHWG18},
studying the cause of the present fast decrease of the dipole field,
which could be a precursor for a reversal,
\citep[see, e.g.,][]{Hulot:2002,finlay2016gyre},
and computational modeling \citep[see, e.g.,][]{olson2009dipole}.
Besides these efforts, 
no consensus has been reached as to what a reliable precursor for a reversal is
\citep[see, e.g.,][]{Constable:2006, Laj:2015}.
This is caused, at least in part, 
by the fact that simulations and the paleomagnetic record
indicate that the details of dipole reversals vary greatly 
\citep[see, e.g.,][]{HFCOM10, glatzmaier2015magnetic}, even
if their directional behavior, as recorded by lava flows, shows some
    degree of similarity from one reversal to the next \citep{VFCH-B12}. 

Here, we re-visit these issues and specifically test the often suggested
possibility that a small value of the dipole's strength could be used 
as a natural indicator of an upcoming dipole reversal.
Our predictions do not distinguish between reversals
and major excursions that lead to a near or total collapse of the axial dipole field,
but end up with the axial dipole rebuilding with the same polarity (see below for why).  
In this study, we therefore collectively refer to reversals or such excursions
as ``low-dipole events'' 
(see Section~\ref{sec:FindingThresholds} for a precise definition of a low-dipole event).

The idea is as follows.
During a low-dipole event,
the dipole intensity
drops to a very low value.
Since the intensity is a continuous function,
it must have approached this low intensity level continuously.
One may thus ask:
can one identify a threshold with the property that if this threshold is passed,
the intensity will continue to decay and a low-dipole event will occur
during a specified time interval, called the prediction horizon.
The prediction horizon is critical to the usefulness of the prediction strategy.
A prediction horizon of several million years, for example, is not useful,
because a low-dipole event is likely to occur over these timescales.
Similarly, a prediction horizon of a few hundred years is not useful
because the low-dipole event may be already in full-swing when we catch it.
Given that it takes several kyr for a dipole reversal
%or excursion
to take place,
a useful prediction horizon should be at least several kyr long.

We can now state the question we want to address more precisely:
\begin{center}
\emph{Can we identify a threshold that is useful for predicting low-dipole events}?
\end{center}
\noindent
We study this question via a hierarchy of models,
ranging from simplified, low-dimensional models, 
to 3D simulations of Earth's magnetic field.
We identify, for each model, a threshold 
by maximizing a skill score that quantifies the skill of the prediction.
When identifying a threshold one should keep in mind that the event 
``a low-dipole event will occur during the prediction horizon'' is rare
in comparison to the event ``no low-dipole event will occur during the prediction horizon''
(at least for useful prediction horizons);
this is addressed by using well-established skill scores that
are robust to imbalances of the occurrence of one event over another.
We carefully discuss the numerical robustness of our approach
and also study robustness with respect to the duration of the training data that
are used to identify a threshold.
We then apply the same methodology to paleomagnetic reconstructions
and discuss the geophysical implications of our study.

Overall, we introduce a new prediction strategy, apply it to four models and two 
paleomagnetic reconstructions (PADM2M and Sint-2000),
and test it with a variety of skill scores.
This causes us to use a large number of acronyms,
most of which are listed in a Table~\ref{tab:AccronymTable} in the Appendix.

\section{Background: model hierarchy and skill scores}
We briefly describe the geomagnetic models we use
and then outline how to assess prediction strategies
via skill scores and receiver operator characteristic (ROC) curves.
Readers who are familiar with the models we use
or with skill scores and ROC curves
may skip this part of the paper.

\subsection{Numerical modeling of the geomagnetic field}
A realistic model for the Earth's magnetic field is a three-dimensional
magneto-hydrodynamic (MHD) model.
Today's MHD models are realistic representations of Earth's magnetic field 
over a large range of spatial and temporal scales \citep{schaeffer2017turbulent,A19,WS19},
but the simulation of dipole reversals
%or excursions
remains a computational challenge
and the number of MHD simulations that exhibit reversals remains limited
\citep{Lhuillier2013,olson2013controls}.
The reason is that Earth-like, high-resolution simulations of the field 
are difficult to do, even with today's super-computers.
As a result, simulations that exhibit reversals often require that 
they be pushed away from the Earth-like regime.
For example, the Ekman number is a control parameter which expresses  
the ratio of the rotation time scale to the viscous time scale. 
Increasing the Ekman number amounts to increasing the kinematic diffusivity
of the fluid and thereby the laminar character of the simulated flow. 
This  in turn decreases the required resolution and the time-to-solution.
For this reason, many reversing simulations are characterized 
by an Ekman number that is much larger than the Ekman number of the Earth's dynamo.

An alternative to 3D simulations are low-dimensional models.
The terminology is perhaps confusing here because the word ``dimensional'' does not
refer to the spatial dimension, but the number of variables within the model.
In this terminology, a 3D model is high-dimensional because it contains a large number of variables
that describe the three-dimensional structure of the fluid flow and its interactions with the magnetic field.
The 3D model we consider below has more than three million variables and, hence,
its dimension is $O(10^6)$.
Low-dimensional models aim to represent selected aspects of the geodynamo 
-- in our case the axial dipole over Myr time scales -- 
with only a small number of variables.
The models we consider have one or three variables
and, hence, dimension one or three -- six orders of magnitude less than the 3D model.
Examples of low-dimensional models include
scalar stochastic differential equations (SDE)
that model the time evolution of the axial dipole as a particle in a double well potential
\citep{Hoyng2001,Schmitt2001,Buffett2013, Buffett2014,Buffett2015,Buffett2015b,MW16, MB19},
scalar SDE's that are inspired by MHD \citep{PFDV09},
and systems of chaotic differential equations that model the interaction of the dipole
with the non-dipole (quadrupole) field, coupled and perturbed by a velocity variable \citep{G12}.

\subsection{The model hierarchy}
We consider three low-dimensional models and one 3D simulation.
We give a concise description of all four models we use
and refer to the original works for further information.
The 3D model we use is unpublished and, for that reason,
we provide more information about the 3D model than the simpler models.

\subsubsection{The deterministic G12 model}
Following \cite{G12}, we consider the ordinary differential equations
\begin{linenomath}\begin{equation}
  \frac{\text{d}Q}{\text{d}t} = \mu Q-VD, \quad 
  \frac{\text{d}D}{\text{d}t} =-\nu D+VQ, \quad
  \frac{\text{d}V}{\text{d}t} = \Gamma-V+QD, 
\end{equation}\end{linenomath}
where $\mu = 0.119$, $\nu = 0.1$, and $\Gamma=0.9$.
Here, $D$ is the dipole
and the variable $Q$ represents the quadrupole
or, more generally, the non-dipole field;
$V$ is a velocity variable that couples $D$ and $Q$.
A change in the sign of $D$ corresponds to a dipole reversal.
We refer to this model as the G12 model.
A typical simulation with G12 is shown in Figure~\ref{fig:Models}.
Here, model time $t$ is scaled to represent the G12 millennium time scale
(1 dimensionless time unit = 4 kyr), see \cite{MFH17}.
The simulation is done by discretizing the differential equation by a fourth-order Runge-Kutta scheme 
(Matlab's ode45).% with a time step of 1 kyr (millennium time scale).

\subsubsection{The stochastic P09 model}
\cite{PFDV09} derived a model for dipole reversals
by considering the interaction of two modes.
%$B(r,t)=a(t)B_1(r) + b(t) B_2(r)$.
Using the symmetry of the equations of magnetohydrodynamics $B\to -B$
in an amplitude equation,
and by assuming that the amplitude has a shorter time scale than a phase,
a stochastic differential equation (SDE) of the form
\begin{linenomath}\begin{equation}
\label{eq:SDE}
	\text{d}x = f(x)\text{d}t + \sqrt{2q}\,\text{d}W,
\end{equation}\end{linenomath}
is derived for the phase, $x$,
where $f(x)$ and $q$ are defined below.
In Equation~\EqRef{eq:SDE}, $W$ is Brownian motion,
a stochastic process with the following properties:
(i) $W(0)=0$; (ii) $W(t)-W(t+\Delta t)\sim\mathcal{N}(0,\Delta t)$;
and (iii) $W(t)$ is almost surely continuous for all $t\geq 0$,
see, e.g., \cite{ChorinHald2013}.
Here and below, $\mathcal{N}(m,\sigma^2)$
denotes a Gaussian random variable with mean $m$,
standard deviation $\sigma$, and variance $\sigma^2$.

More specifically, the SDE for the phase is defined by
\begin{linenomath}\begin{equation}
\label{eq:P09}
	f(x)=\alpha_0+\alpha_1\,\sin(2x),\quad \sqrt{2q} = 0.2\sqrt{\left\vert\alpha_1\right\vert}.
\end{equation}\end{linenomath}
We use the same parameters as in \cite{PFDV09},
$\alpha_1=-185\,\mbox{Myr}^{-1}$, $\alpha_0/\alpha_1 = -0.9$. 
The dipole, $D$, can be calculated from the phase by $D = R\cos(x+x_0)$.
Following \cite{MFH17}, we set $x_0=0.3$ and $R=1.3$
(the latter scales the dipole variable $D$ to have approximately the same time average
as the relative paleointensity reported by the reconstruction of Sint-2000
\citep{SINT2000}).
For the reminder of this paper, we refer to this model as the P09 model.

Note that the parameters define the model's time scale.
The parameters are chosen so that the P09 model exhibits reversals and excursions,
and so that its reversal rate is comparable to that of Earth's dipole.
This is illustrated in Figure~\ref{fig:Models},
where a typical simulation result with this model is shown.
For a simulation we discretize the differential equation using a forward Euler-Maruyama method \citep{KloedenPlaten}.
The time step is 1 kyr.

\subsubsection{The double well model}
A simple model for reversals of a quantity 
(not necessarily Earth's dipole field) 
is a particle in a double well potential.
Such a model is defined by an SDE model as in equation~(\ref{eq:SDE}),
and with an $f(x)$  that is equal to the negative gradient of a double well potential.
Variations of this model for geomagnetic dipole reversals have been considered by many researchers 
\citep{Hoyng2001,Schmitt2001,Buffett2013, Buffett2014,Buffett2015,Buffett2015b,MW16, MB19}.
The basic idea is that the state, $x$, of the SDE is within one of the two wells
of the double well potential and is pushed around by noise 
(the Brownian motion $\sqrt{2q}\,\text{d}W$).
When the noise builds up towards one side of the well,
the state may cross over to the other potential well.
One can identify a transition from one well to the other as a reversal of Earth's dipole.

We use a recent version of this model, called the Myr model in \cite{MB19},
for which
\begin{linenomath}\begin{equation}
        f(x) = \gamma\frac{x}{\bar{x}}\cdot
        		\left\{\begin{array}{l}
        			(\bar{x}-x), \quad \text{if } x\geq 0\\ 
		         (x+\bar{x}), \quad \text{if } x< 0
		         \end{array}  \right.,
\end{equation}\end{linenomath}
where $\gamma=0.1 \text{ kyr}^{-1}$, $\bar{x}=5.23 \cdot10^{22}\text{ Am}^2 $
and $q=0.34 \cdot10^{44}$ A$^2$m$^4$ kyr$^{-1}$.
These parameters define the model's natural time scale
and the values we chose are based on configuration (a) in \cite{MB19},
which implies that the model's reversal rate is comparable with Earth's reversal rate.
For the reminder of this paper, we refer to this model as the DW model.
A typical simulation of this model is shown in Figure~\ref{fig:Models}.
For a simulation we discretize the equation using a  fourth-order Runge-Kutta for the deterministic part
and a forward Euler-Maruyama for the stochastic component \citep{KloedenPlaten}.
The time step is 1 kyr.

\subsubsection{The 3D model}
\label{sec:3d_pres}
We consider a three-dimensional, convection-driven, dynamo 
simulation which exhibits polarity reversals and dipole excursions.
The simulation we consider has not yet been published
and is part of an ensemble of reversing simulations run by 
N.~Schaeffer (ISTerre, CNRS, Universit\'e Grenoble Alpes),
 A.~Fournier and T.~Gastine (both affiliated with
 Universit\'e de Paris, Institut de Physique du Globe de Paris). 
We refer to this simulation simply as the 3D model for the rest of this paper.

The 3D model uses
a pseudo-spectral approximation to solve the set of equations governing
rotating dynamo action in a spherical shell geometry 
\citep[see, e.g.,][for details]{Christensen2015}. 
The scales chosen to non-dimensionalize the set of equations are the same as those used 
by e.g. \cite{schaeffer2017turbulent}. 
The radius ratio of the inner-core
boundary to the core-mantle boundary is set to its present-day value. 
The 3D model has no-slip boundary conditions on the inner-core  
and core-mantle boundaries, and it assumes that the inner core is conducting. 
The four non-dimensional control parameters, as defined e.g. in \cite{schaeffer2017turbulent},
are as follows. The Ekman number is~$10^{-4}$, the Prandtl number is~$1$, the 
magnetic Prandtl number is~$3$ and the Rayleigh number is~$15000$.
These choices result in an average hydrodynamic Reynolds number of $216$
(recall that the hydrodynamic Reynolds number is defined as the product of the
root-mean-squared velocity by the shell thickness divided by the kinematic viscosity).
The open-source, freely available
xshells code \footnote{\htmladdnormallink{https://nschaeff.bitbucket.io/xshells/}
{https://nschaeff.bitbucket.io/xshells/}} 
is used to numerically solve the equations. This code combines the finite difference method in the 
radial direction with a spherical harmonic representation of field variables in the horizontal
direction, using the dedicated SHTns library \citep{schaeffer2013efficient}.
To ensure numerical convergence, a hyperdiffusivity is applied beyond spherical harmonic degree $55$. 

The resolution of the 3D model is 
defined by the triplet $\left(N_r,\lmax,\mmax\right)$, 
giving the number of points used in the radial direction together with the maximum
degree and order used in the horizontal approximation of field variables with spherical harmonics. 
Since  five scalar fields  are discretized, the total number of degrees of freedom of a simulation
is  $ { \mathcal O} \left(5 N_r \lmax \mmax \right)$. 
Namely, the triplet defining the resolution is $(144,79,63)$,
which results in about $3.6 \times 10^6$ variables --
recall that G12 has three variables, P09 and DW have one variable.

As discussed above, time in the 3D model is non-dimensional.
To scale to geophysical time, 
one first computes the non-dimensional secular-variation timescale of the non-dipole field 
up to spherical harmonic degree~13, 
based on the average power spectra of the magnetic field and its secular variation
\citep[see][]{lhuillier2011grl}. The rescaling of the time axis is then 
performed under the assumption that the dynamo simulations and the Earth
share the same secular-variation time scale, equal to $415$~yr. 
With this scaling, the simulation time of the 3D model is $147$~Myr;
the time step is 43.09 years.
The number of reversals that occur during this time frame is $109$. 

In contrast to the other models described above, 
its 3D nature makes this dynamo model amenable to quantitative comparisons
against more observed properties of the Earth's magnetic field than just the axial dipole.
From a morphological standpoint, 
the 3D model produces a magnetic 
 field whose large-scale properties at the core-mantle boundary
 are in ``good'' agreement with well-established observations,
according to the four criteria introduced by  \cite{Christensen2010}: 
\begin{enumerate}[(i)]
\item 
the axial dipole to non-axial dipole energetic ratio;% (model average: 0.72)
\item 
the equatorially symmetric to antisymmetric non-dipole energetic ratio;% (model average: 1.36)
\item 
the zonal to non-zonal energetic ratio;% (model average: 0.18)
\item 
the flux concentration factor.% (model average: 2.45)
\end{enumerate}
The terrestrial reference values for these four quantities are
respectively (1.4, 1.0, 0.15, 1.5) \citep{Christensen2010}.
For the 3D model, we compute average values of these quantities of,
respectively (0.72, 1.36, 0.18, 2.45).
This leads to an average misfit $\chi^2$ of 1.94, while  
the median value of $\chi^2$ over the course of the numerical integration is $2.89$.
We refer to \cite{Christensen2010} for further details. 

% I suggest we start a new paragraph here, in order to give the reader a break
From a paleomagnetic perspective, it is worth noting that 
\cite{sprain2019assessment} recently introduced a method
to assess the degree of spatial and temporal agreement of a simulated
dynamo field with the long-term ($\sim10$ Myr) paleomagnetic field.
The agreement is defined on the basis of five  properties of the
paleomagnetic field, namely the inclination anomaly, the virtual
geomagnetic pole dispersion at the equator, the latitudinal
variation in virtual geomagnetic pole dispersion, the normalized
width of virtual dipole moment (VDM) distribution, and the dipole field
reversals (in terms of the relative time spent by the dipole
at transitional latitudes lower than $45^\circ$).
This quantity, termed $\Delta Q_{PM}$, is the sum of five misfits,
one for each criterion. 
For the 3D model, 
we find the following values of the misfit  for each criterion
\begin{linenomath}\begin{equation*}
\begin{aligned}
\Delta Q_{PM} (\mbox{inclination anomaly} ) & = &0.84, \\ % 0.83
\Delta Q_{PM} (\mbox{equatorial dispersion} ) & = & 0.39, \\% 1.67,\\ 
\Delta Q_{PM} (\mbox{latitudinal dispersion} ) & = & 0.95, \\%0.07, \\
\Delta Q_{PM} (\mbox{VDM distribution} ) & = & 0.81, \\
\Delta Q_{PM} (\mbox{reversals} ) & = & 1.45,
\end{aligned}
\end{equation*}\end{linenomath}
for a total $\Delta Q_{PM}=4.43$. %83$. 
This value is good, according to \cite{sprain2019assessment}, 
who argue that individual misfits lower 
than unity indicate an adequate similarity with the paleomagnetic field.
For the study of interest here, we note that the width of the VDM distribution is adequately captured by
the 3D model. On the other hand, the simulated dipole spends a fraction of time at transitional latitudes ($1.2$\% of
the model integration time)  smaller than what is inferred for Earth 
over the last 10 Myr, which is expected to lie somewhere between $3.75$\%                 
and $15.0$\% \citep[see][for details]{sprain2019assessment}. 
In summary, based on a series of metrics that have come to the fore,  this 3D model 
compares favorably against the recent and more ancient geomagnetic field. 

\subsection{Similarities and differences across the model hierarchy}
\begin{figure}%%[tb]
	\centering
	\includegraphics[width=1\textwidth]{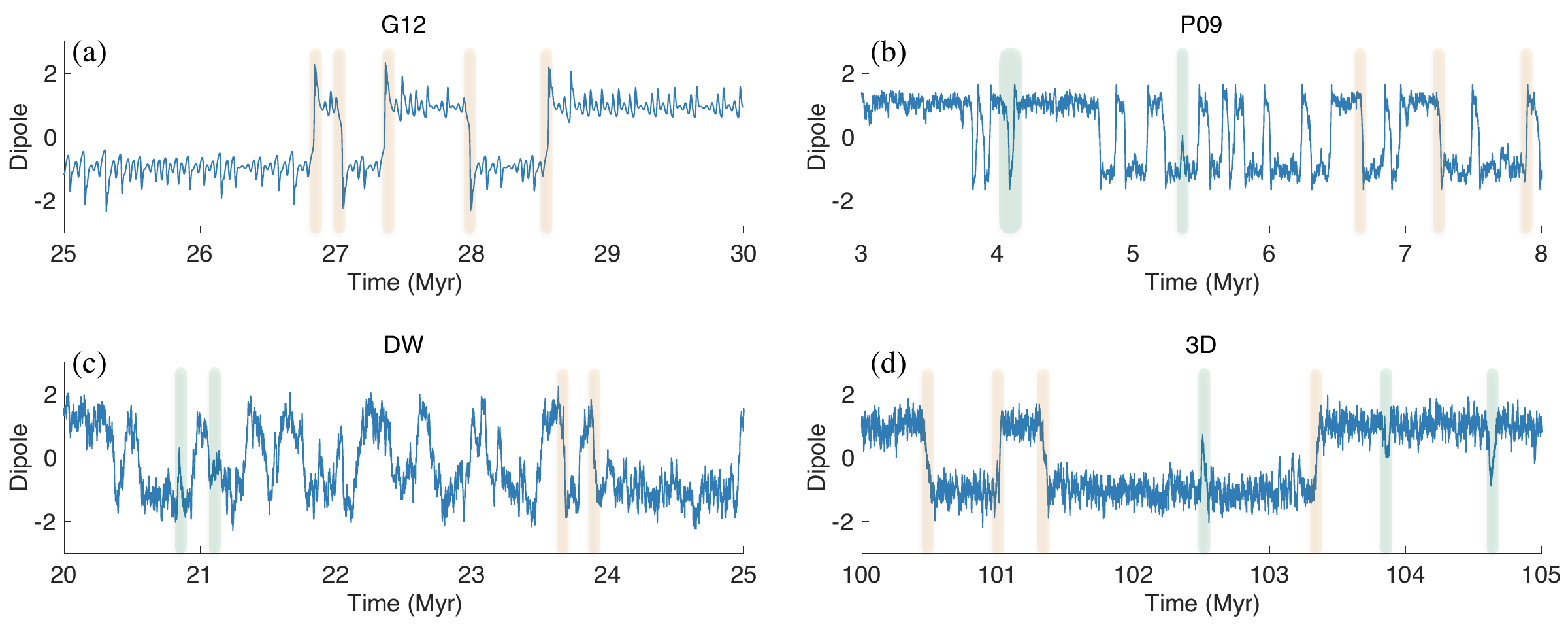}
	\caption{Signed dipole as a function of time for the four models considered in this study. 
	(a) G12, (b) P09, (c) DW and (d) 3D.
	In each case, the amplitude is scaled so that the average absolute value of the time series is one.
	Some reversals and excursions are highlighted in light red and light blue.}
	\label{fig:Models}
\end{figure}

\begin{figure}%%[tb]
	\centering
	\includegraphics[width=1\textwidth]{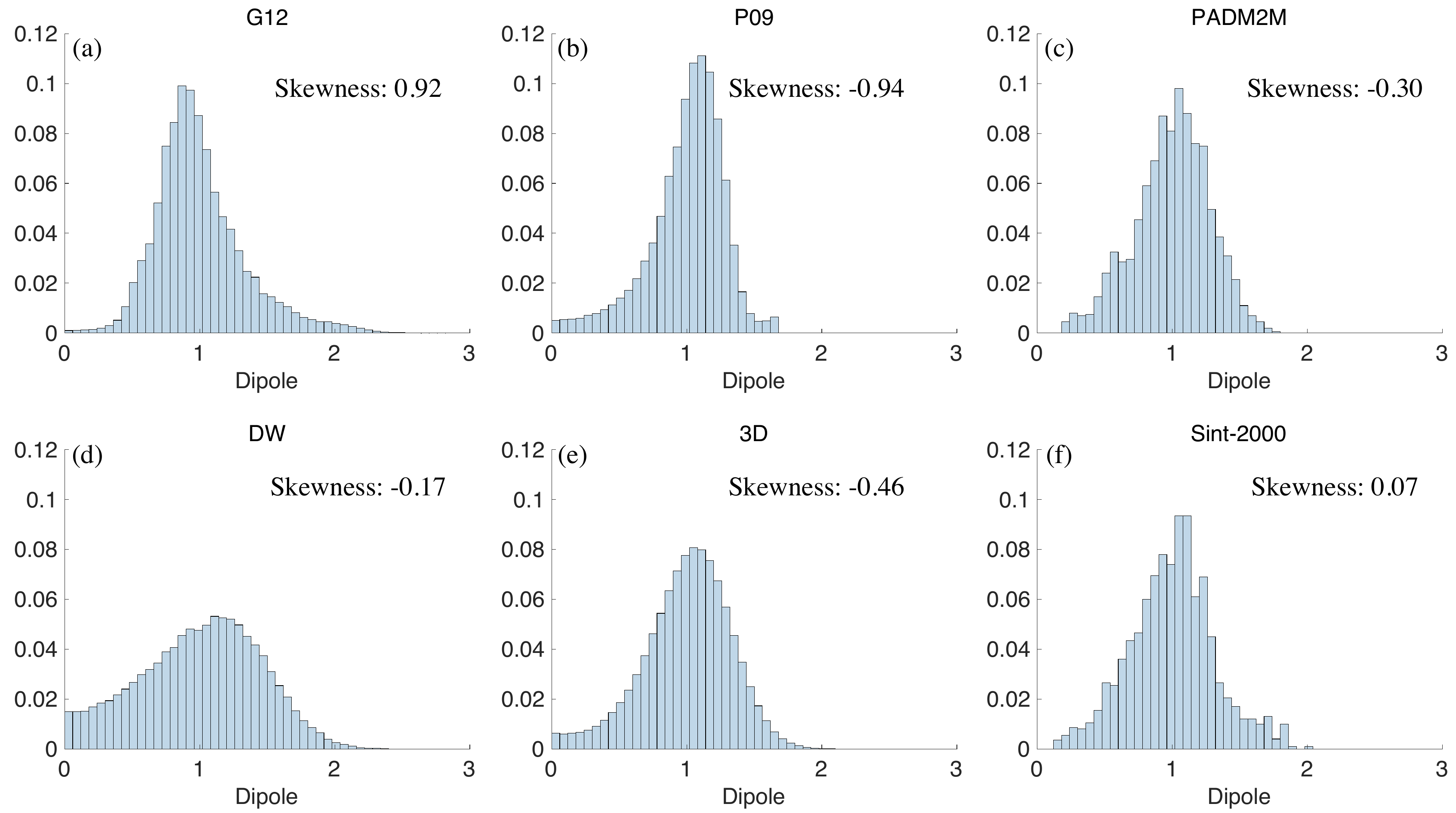}
	\caption{Scaled histograms of the dipole intensities of the four models
	and two paleomagnetic reconstructions (PADM2M and Sint-2000).
	(a) G12, (b) P09, (c) PADM2M, (d) DW, (e) 3D, and (f) Sint-2000.
	The y-axes of all histograms are scaled so that the area under the graph is equal to one
	and the x-axes are scaled so that one corresponds to the average intensity.
	Also shown is the skewness which indicates the degree of asymmetry
	in the distribution.
	A thicker tail near zero suggests that a model, or reconstruction,
	lingers in a state of low intensity.}
	\label{fig:AmpHists}
\end{figure}

Figure~\ref{fig:Models} shows the axial dipole as a function of time
for each model, scaled so that the average absolute value of the time series is one,
and with the sign indicating polarity:
a negative sign indicates today's polarity,
a positive sign corresponds to a reversed polarity.
 The figure shows the evolution of the models' dipoles on their natural time scales described above.
Each model exhibits reversals and excursions
as observed in paleomagnetic reconstructions,
e.g., PADM2M and Sint-2000
\citep{PADM2M,SINT2000},
but these events occur over  different time scales
and, for some of the models,
the events also occur over time scales that are different from what is observed in 
Earth's axial dipole field.
The 3D model, for example, 
has a reversal rate of about 0.7 reversals per Myr,
but the reversal rates of the DW and P09 models are about 5 reversals per Myr,
which is comparable to the average reversal rate of Earth over the last 25 Myr \citep{O12}.
In addition, 
the way a reversal occurs in each model can be different.
Reversals of the G12 model
are characterized by a continuous decay in intensity,
immediately followed by a rapid increase in intensity.
The  DW lingers at low intensity longer than the other models
(see also Figure~\ref{fig:AmpHists}).
We also observe that the DW and 3D models,
and to a lesser extent the P09 model, 
exhibit multiple, rapid fluctuations in sign during a reversal,
see Figure~\ref{fig:3D_Illu}.
This behavior is not observed in the G12 model.

Differences between the various models can be illustrated further by  
comparing histograms of their intensities,
shown in Figure~\ref{fig:AmpHists},
which also provides the models' Pearson's moment coefficient of skewness 
(third standardized moment, \cite{KK62})  for the four models.
It is clear that all models, except G12, are characterized by a negative skew.
Thus, the P09, DW and 3D models spend more time at a lower than average intensity than
at a higher than average intensity.
The G12 model tends to spend more time at a higher than average intensity than
at a lower than average intensity.
The DW model has the smallest skew (in amplitude) and a thicker left tail than the other three models,
which indicates that it spends a considerable amount of time in low-intensity states
(but other formulations of double-well models, with different parameters
or even different parameterizations of the potential,
may behave differently).
Also shown in Figure~\ref{fig:AmpHists}
are histograms of the intensities of two paleomagnetic reconstructions,
PADM2M and Sint-2000 \citep{PADM2M,SINT2000},
which document the time evolution of the virtual axial dipole moment (VADM)
over the past 2 Myr at a frequency of 1 per kyr.
PADM2M and Sint-2000 thus contain 2000 points
and the histograms are not as well resolved as those of the four models
(for which we used substantially longer simulations).
This is also evident from the difference in the skewness,
which is positive for Sint-2000, but negative for PADM2M.
These values should be used with caution,
because the estimates of skewness are contaminated by large sampling error
(based on only 2000 intensities)
and by the fact that low intensities (below 10\%) are not present in these reconstructions.
In view of the large uncertainty in the reconstructions,
all four models are reasonable,
at least qualitatively, i.e., in view of Figures~\ref{fig:Models} and~\ref{fig:AmpHists},
although all four models are constructed from drastically different assumptions
and with very different modeling goals in mind.

\subsection{Predictions, skill scores, and ROC curves}
\label{sec:PredictionSkills}
We want to predict whether a low-dipole event
will occur within an a priori specified time interval,
called the prediction horizon.
We thus consider only two outcomes of an experiment.
Outcome 1: yes, the event occurred during the prediction horizon;
outcome 2: no, the event did not occur during the prediction horizon.
As is common, we denote the outcomes of an experiment by ``positives'' and ``negatives:''
\begin{linenomath}\begin{align*}
\text{Positive (P):} & \text{ the event occurred.}\\
\text{Negative (N):} & \text{ the event did not occur.}
\end{align*}\end{linenomath}
With two possible outcomes of an experiment, a
prediction can result in one of four possibilities:
\begin{linenomath}\begin{align*}
\text{True positive (TP):} & \text{ predict  that an event will occur and the event occurs.}\\
\text{False positive (FP):} & \text{ predict that an event will occur, but the event does not occur.}\\
\text{True negative (TN):} & \text{ predict that an event will not occur and the event does not occur.}\\
\text{False negative (FN):} & \text{ predict that an event will not occur, but the event occurs.}
\end{align*}\end{linenomath}
The concepts and ideas described here have been used in many areas.
We make an effort to be consistent in the terminology and to bring up only the definitions we need,
sticking to commonly used names \citep{F06,CJ20,Joliffe15}.
For a more thorough review of the such predictions in the context of (medical) imaging, see \cite{Barrett},
Chapter~13,
where, a prediction strategy of the type discussed here is called a binary decision.
In the language of machine learning,
the problem of predicting ``an event will occur within the horizon''
or ``no event will occur within the horizon''
is called a classification problem,
similar to distinguishing dogs from cats \citep{GBC2016}.

It is clear that a good prediction strategy should be characterized by 
a large number of true positives and true negatives,
but a small number of false positives or false negatives.
A skill score is a quantitative means for describing the quality of a prediction strategy.
There is a large number of skill scores
and these usually require that one applies the prediction strategy, say $n$ times,
followed by counting the number $\Pos$ of positives that occurred, 
the number $\Neg$ of negatives that occurred,
and the true\slash false positives and true\slash false negatives.
Which of the many skill scores is most appropriate depends on the problem one wishes to solve.
For example, one may define the accuracy by
\begin{linenomath}\begin{equation}
\label{eq:ACC}
 \ACC = \frac{\TP+\TN}{\Pos+\Neg}.
\end{equation}\end{linenomath}
A good prediction strategy should be characterized by a high accuracy,
but a bad prediction strategy may also be characterized by a high accuracy.
For example, the event ``a low-dipole event occurs within the prediction horizon'' is rare
compared to the event ``no low-dipole event occurs within the prediction horizon''
(unless the prediction horizon is large).
This means that the prediction strategy ``predict that no low-dipole event occurs within the prediction horizon''
is characterized by a high accuracy,
but this strategy is useless because it can never achieve a true positive
(the event ``a low-dipole event occurs within the horizon'' is never predicted).
Other skill scores, e.g., the $\F$ score
\begin{linenomath}\begin{equation}
\label{eq:F1}
	\F = \frac{2\,\TP}{2\TP+\FP+\FN},
\end{equation}\end{linenomath}
the critical success index (CSI)
\begin{linenomath}\begin{equation}
\label{eq:CSI}
	\text{CSI} = \frac{\TP}{\TP+\FP+\FN},
\end{equation}\end{linenomath}
or Matthew's correlation coefficient (MCC)
\begin{linenomath}\begin{equation}
\label{eq:MCC}
 \MCC = \frac{\TP \cdot \TN - \FP \cdot \FN}
 {\sqrt{(\TP+\FP)(\TP+\FN)(\TN+\FP)(\TN+\FN)}},
\end{equation}\end{linenomath} 
are designed to alleviate these issues
and are applicable in problems where the occurrence of the event is rare.

One can also compute the true-positive-rate (TPR) and false-positive-rate (FPR),
defined by:
\begin{linenomath}\begin{equation}
\label{eq:TPRFPR}
\TPR =\frac{\TP}{\Pos},\quad
\FPR =\frac{\FP}{\Neg}.
\end{equation}\end{linenomath}
TPR and FPR have the desirable property that they are independent of the 
frequency of the event and a good prediction strategy should have a 
high TPR and low FPR,
independently of how often an event occurs.
Ideally, $\TPR=1$, so that all events are predicted correctly,
and $\FPR =0$, so that no false positives occur.

\begin{figure}%[tb]
	\centering
	\includegraphics[width=0.5\textwidth]{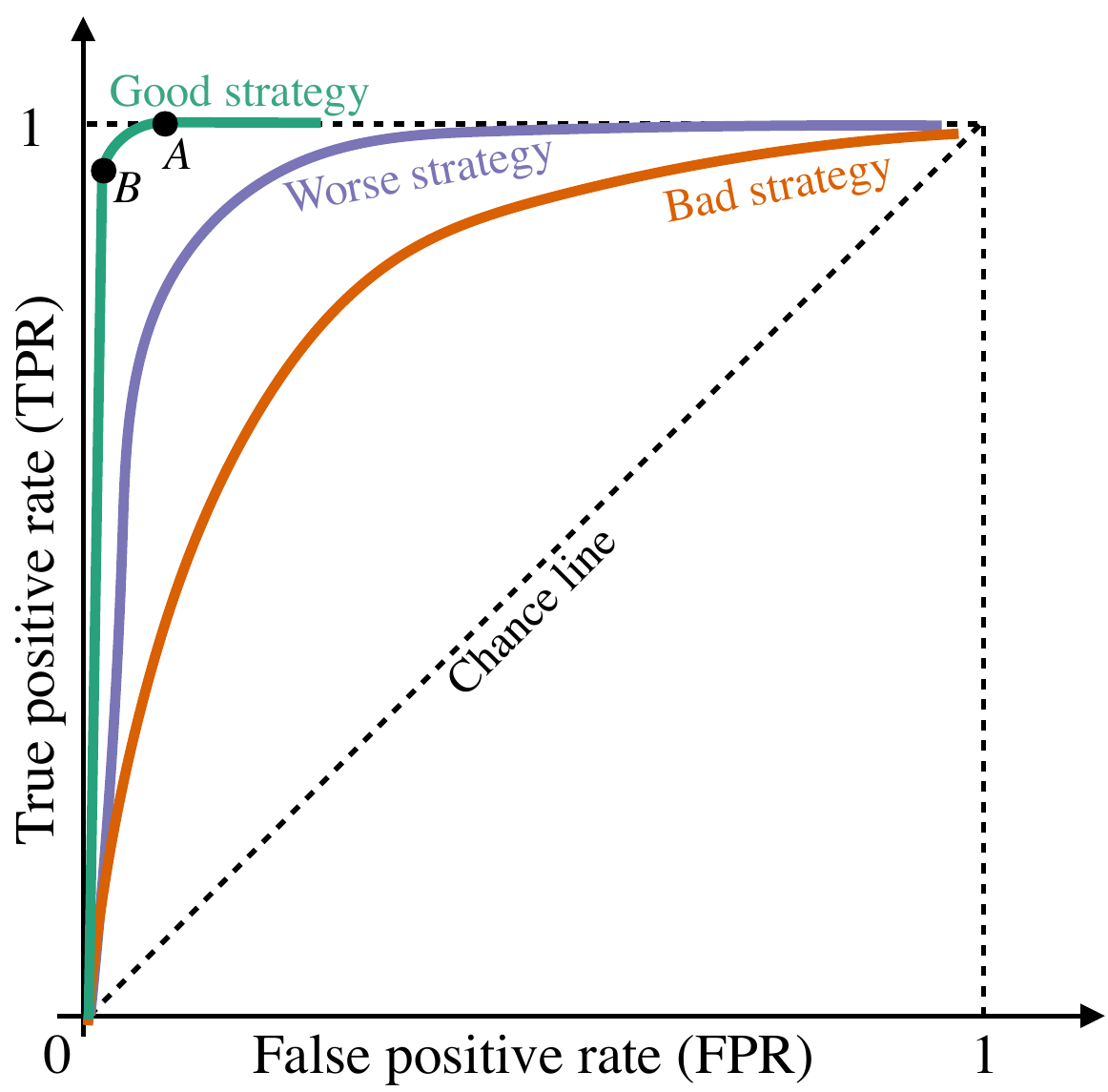}
	\caption{Three examples of ROC curves.
	Two threshold levels, labeled $A$ and $B$, are identified on one of the curves
	(see text for the definition of the chance line).}
	\label{fig:ROCIllu}
\end{figure}

The bulk of this paper is concerned with predictions 
based on whether the dipole is below a specified threshold. 
Naturally, one may investigate how the skill of the predictions depends on the threshold.
For example, one can compute skill scores 
as functions of a varying threshold
and determine an optimal threshold as the one that maximizes the skill score.
One can also compute the $\TPR$ and $\FPR$ as functions of the threshold.
The line that a varying threshold traces out in $\TPR-\FPR$ space
is called the receiver operating characteristic (ROC) curve.
Three examples of ROC curves are illustrated in Figure~\ref{fig:ROCIllu}.
The figure also shows the chance line,
defined by a straight line at a $45^{\circ}$ angle,
on which the (FTR,TPR) points should lie when randomly guessing occurence of events,
and varying the probability with which one makes this guess.

The ROC curve of a good prediction strategy should be above the chance line
and should quickly transition from the origin towards $(0,1)$,
thus being characterized by a high TPR and a small FPR. 
One can use ROC curves as qualitative tools to assess different prediction strategies.
We have included labels for the three ROC curves in Figure~\ref{fig:ROCIllu}, 
that identify which strategies are good, worse or bad.

We note that, besides an impressively large body of work across many disciplines,
it remains difficult to unambiguously argue that a prediction strategy is good or bad,
or if one prediction strategy is better than another.
As a simple example, consider the green ROC curve in Figure~\ref{fig:ROCIllu}
with threshold levels labeled by $A$ and $B$.
It is not easy to say which threshold level one should choose.
Threshold $A$ leads to the smallest $\FPR$ while also achieving $\TPR=1$,
but threshold $B$ achieves a smaller $\FPR$ than threshold $A$,
at the cost of a slightly smaller $\TPR$.
The difficulties arise because many issues, 
such as how dangerous false positives are compared to false negatives,
are problem dependent and remain subjective.

\section{Finding thresholds for the prediction of low-dipole events}
\label{sec:FindingThresholds}
\begin{figure}%[tb]
	\centering
	\includegraphics[width=1\textwidth]{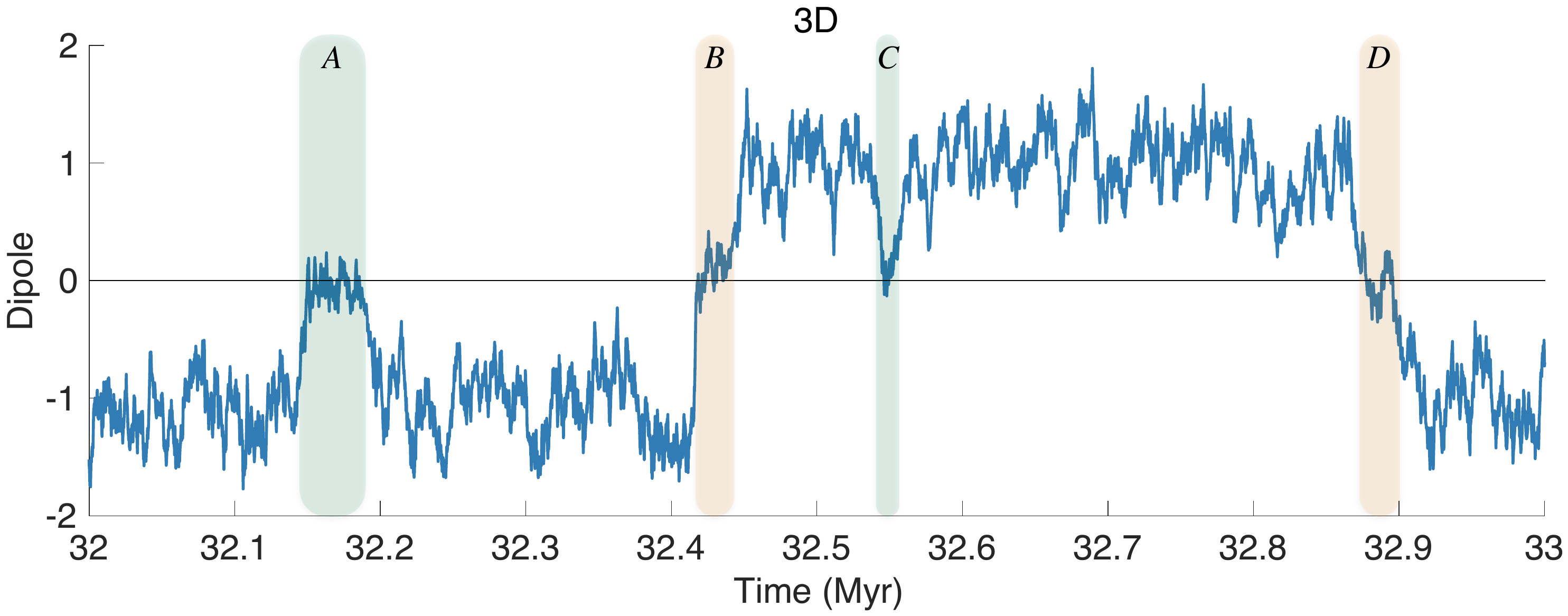}
	\caption{Excerpt of the 3D simulation
	showing the signed dipole as a function of time (same scaling as in Figure~\ref{fig:Models}).
	Four events are labeled $A-D$.}
	\label{fig:3D_Illu}
\end{figure}

In simple terms, our prediction strategy is
\noindent
\begin{center}
\emph{If the dipole intensity drops below a threshold, 
then the field will continue to decay and a low-dipole event will occur within the prediction horizon.}
\end{center}
The situation, however, is more delicate because the models 
exhibit complex behavior while undergoing reversals or major excursions.
% or excursions.
The subtleties can be illustrated by 
considering the excerpt of the 3D simulation shown in Figure~\ref{fig:3D_Illu},
where we highlight
reversals and major excursions during which the axial dipole field temporarily changed sign.
%several reversals and excursions,
%labeled as events $A-D$.
One wishes to define a reversal event as the transition of 
a \textit{strong} dipole field in one polarity,
to a \textit{strong}  dipole field in the opposite polarity,
rather than just a short-term temporary change of polarity. 
%rather than a change of polarity.
In fact, the field may quickly change polarity several times while the field is weak
when undergoing a reversal.
This occurs in the 3D simulation during the events labeled $B$ and $D$ in Figure~\ref{fig:3D_Illu}.
Similarly, one wishes to interpret event $A$ (or $C$)
as a single major excursion, rather than a sequence of reversals.
We thus revise the simple prediction strategy above to ensure that
each reversal or major excursion,
labeled by $A$ -- $D$ in Figure~\ref{fig:3D_Illu}, 
is considered as a single low-dipole event.
%We further do not distinguish between reversals or excursions in our predictions
%and call reversals and excursions collectively low-dipole events.
A careful definition of low-dipole events is provided below.

\subsection{Precise formulation of threshold-based predictions}
\label{sec:Definitions}
We introduce definitions that allow us to clearly specify the events and predictions whose
skill we want to study.
We start with the definition of the event we want to predict.

\vspace{2mm}\noindent
{\bf{Definition: }}\emph{Low-dipole event}.
A low-dipole event starts when the intensity drops below a specified value, called the \emph{start-of-event threshold} (ST), 
or if a the dipole changes its sign\footnote{The addition of the ``or-statement'' is relevant only in the context of paleomagnetic reconstructions, see Section~\ref{sec:Application}.},
and ends when the intensity exceeds a second specified value, called the \emph{end-of-event threshold} (ET).
The \emph{event duration} is the time interval from start to end of the event.

\vspace{2mm}\noindent
This definition ensures that a low-dipole event describes reversals and major excursions,
because the field may drop below the ST, but can build back up above the ET without changing polarity.
Events $A$ and $C$ in Figure~\ref{fig:3D_Illu} are examples of this situation.
We also emphasize that the event duration is defined implicitly
by the start and end of an event and may vary considerably across several low-dipole events.
In Figure~\ref{fig:3D_Illu}, for example, 
the low-dipole event $A$ has a much larger event duration than event $C$,
but both events are major excursions.

To define a strategy for predicting low-dipole events, 
we introduce the \emph{prediction horizon} (PH),
which is the time window during which we predict that a low-dipole event will start to occur.
Note that we do not make any prediction as to when precisely the low-dipole event starts -- 
we merely predict that a low-dipole event will start (or not) at some point during the PH.
We further make no predictions as to when the low-dipole event will end.
%In \cite{MFH17}, predictions of this type are called coarse predictions.
With the above definitions, the prediction strategy can be stated precisely.

\vspace{2mm}\noindent
{\bf{Definition: }}\emph{Threshold-based predictions for low-dipole events}.
We predict that a low-dipole event will start to occur within the prediction horizon
if the intensity drops below a \emph{warning threshold} (WT);
we predict that no low-dipole event will start during the prediction horizon if
the intensity is above the WT;
we stop making predictions from the time the low-dipole event started
(intensity below ST) until the event ends (intensity above ET).

\begin{figure}%[tb]
	\centering
	\includegraphics[width=1\textwidth]{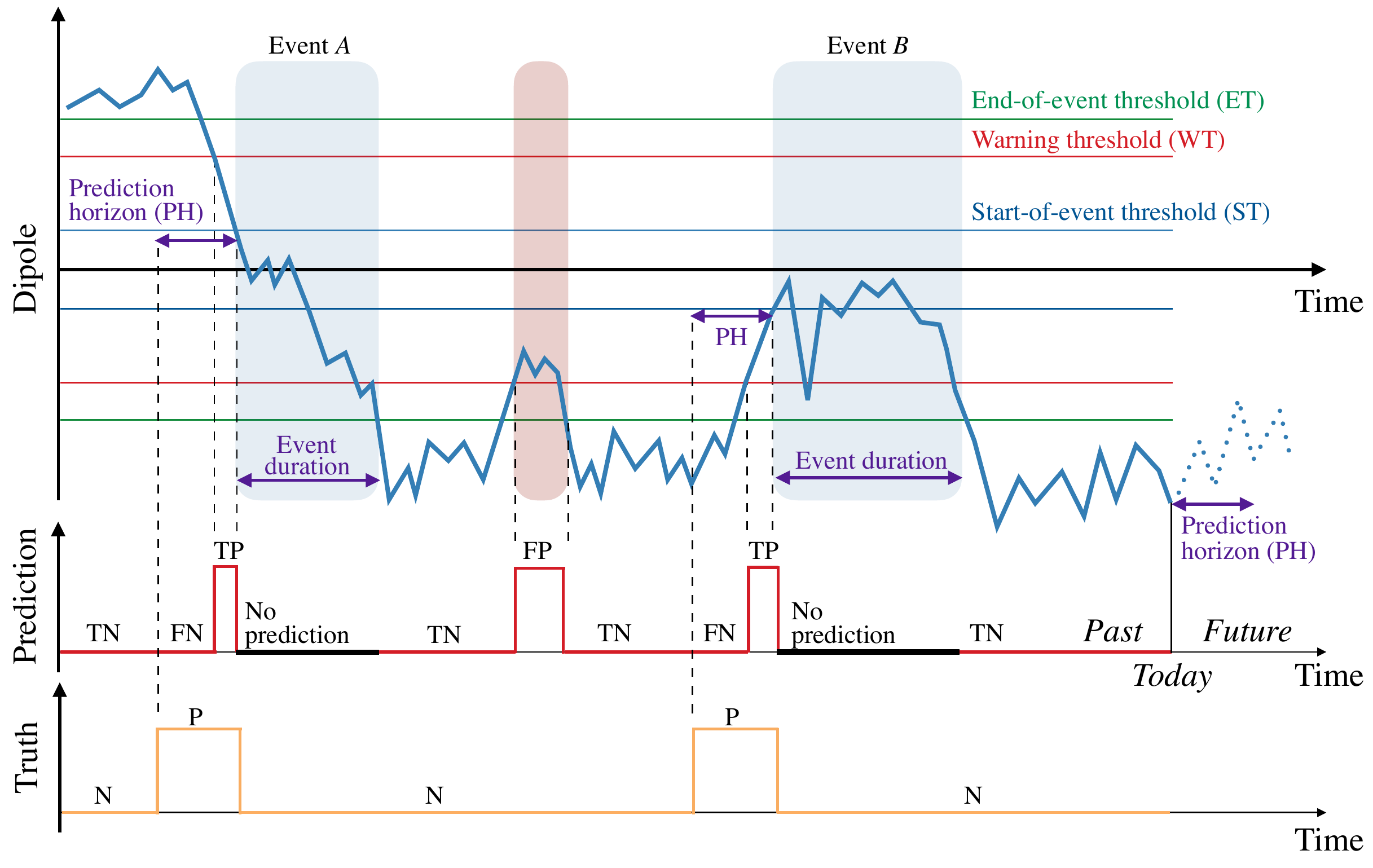}
	\caption{Illustration of the prediction strategy.
	\emph{Top}: dipole (solid blue) as a function of time.
	The thin blue, green and red horizontal lines represent the start-of-event,
	the end-of-event 
	and the warning thresholds.
	Two low-dipole events are labeled $A$ (reversal) and $B$ (excursion),
	and we indicate their event durations.
	Highlighted in red is a period of low intensity,
	which is not a low-dipole event,
	but where the low intensity causes false positives (FP).
	Towards the right, we illustrate a prediction over a given prediction horizon,
	which will lead to true negatives (TN).
	The prediction horizon also defines the true labels, (see bottom panel).
	\emph{Center}: prediction as a function of time.
	The red line at zero corresponds to the prediction ``no low-dipole event occurs during the prediction horizon,''
	and the red line at one corresponds to the prediction ``a low-dipole event occurs during the prediction horizon.''
	The thick black line segments correspond to periods during which no prediction is made.
	For events $A$ and $B$, we first observe TNs,
	followed by false negatives (FN), caused by the warning threshold being small;
	then we observe TPs followed by a period during which no prediction is made.
	\emph{Bottom}:
	true occurrences of low-dipole events within the prediction horizon.
	The orange line at zero corresponds to negatives (N), i.e., ``no low-dipole event occurs during the prediction horizon.''
	The orange line at one corresponds to positives (P), i.e., ``a low-dipole event occurs during the prediction horizon.''
	}
	\label{fig:Strategy_Illu}
\end{figure}

\vspace{2mm}\noindent
We make no predictions while the event is observed,
because a prediction made while an event is happening is of limited use.
Our prediction strategy is illustrated in Figure~\ref{fig:Strategy_Illu}.
In this illustration, ST and ET are chosen such that
Event~$A$ is a single event;
fast oscillations in polarity occur while the field is weak.
%No predictions are made from the time the event starts (intensity below ST)
%to the time that the event has ended (intensity above ET).
The prediction strategy leads to TNs followed by FNs and TPs for Events~A and~B.
The false negatives occur because, given the prediction horizon and average intensity,
the warning threshold is small,
so that the events tend to be predicted a little too late.
The figure also illustrates FPs,
which occur when the field drops below the WT, but no low-dipole event occurs
because the field does not continue to drop below the ST.
True negatives occur often,
because reversals and low dipole events are rare.
In the figure, TNs occur whenever the ``Truth'' (bottom) and the prediction (center) are both at zero.

Finally, note that with our definitions,
one may require that 
\begin{linenomath}\begin{equation}
	\text{ST}< \text{WT}< \text{ET},
\end{equation}\end{linenomath}
because other choices for WT lead to rather strange prediction strategies.
If $\text{ET}\leq\text{WT}$, for example,
then a low-dipole event would be predicted 
immediately after an event just ended.

\subsection{Scaling thresholds and re-scaling time}
\label{sec:ScalingTimeAndThresholds}
For each model, we define all three thresholds (warning, start-of-event, and end-of-event thresholds)
as a fraction of the average intensity of the model.
Such a scaling of the thresholds makes comparisons across the hierarchy of models easier to understand.
For the rest of this paper we fix the start-of-event and end-of-event thresholds as: ST =10\% and ET=80\%.
With these choices, we focus on events that start when the intensity is very low 
and which end when the field has nearly fully recovered
(see Figure~\ref{fig:3D_Illu}).
The choice of ST =10\% is guided by the consideration that we want to focus on events that correspond to reversals
and major excursion.
During a reversal, the signed dipole can reach an arbitrarily low value,
before switching sign.
During a major excursion, the dipole amplitude is very low,
but we do not necessarily observe a switch in the sign.
Moreover, paleomagnetic reconstructions, such as PADM2M and Sint-2000 \citep{PADM2M,SINT2000}, 
have difficulties with resolving small dipole values.
The paleomagnetic reconstructions we consider below
consist of signed Virtual Axial Dipole Moments (VADM),
which are proxies for the true axial dipole magnitude.
The weakest VADMs recorded are about 10-20\% of the present axial dipole field intensity  
\citep[see, e.g.,][]{Constable:2006,HFCOM10}.
This is caused by (i) VADM reconstructions sensing the non-dipole field during a low-dipole event;
(ii) VADM reconstructions are temporally filtered by sediment recording processes;
and (iii) additional smoothing is introduced by modeling choices and stacking of the
relative paleointensity (RPI) records 
(some of the individual records may have a higher resolution
and features that are not aligned in time are smoothed out).
As we will see, by choosing ST~=10\%, we ensure that only events that experienced
at least one temporary change of sign in the axial dipole are considered as events of interest
within PADM2M and Sint-2000 (see Section~\ref{sec:Application}).
%----------

Nonetheless, the precise values of ST and ET are not critical
because our overall approach is robust with respect to choices. 
This is evident from a limited number of numerical experiments we performed 
with different choices of ET and ST.
Specifically, we tried the combinations ST=10\% and ET = 50\%,
ST=20\% and ET = 50\%, and ST =20\% and ET=80\%
and obtained qualitatively and quantitatively similar results.

We rescale time in each model so that the prediction results 
are comparable across the hierarchy of models.
A natural choice for this time scale is the \emph{average event duration} (AED).
That is, we compute the average event duration
given the natural time-scale of each model,
and then re-scale time so that one time unit corresponds to one average event duration.
The average event duration (AED) for each of the models is listed in Table~\ref{tab:MainTable}.
For the simplified models (G12, P09 and DW), 
the statistics of the event duration are computed from simulations
that include about 550 events.
For the 3D model, 
we use the entire duration of the simulation to compute the statistics of the event duration.

The prediction horizon is defined as a fraction of the average event duration.
We focus on the prediction horizon $\text{PH}=1\times$ AED,
i.e., we focus on short-term predictions of low-dipole events,
attempting to predict whether a low-dipole event will start
with a lead time comparable to the event's duration.
We also consider prediction horizons of $0.5 \times$ AED or $1.5\times$
AED, to show how the prediction skill
degrades with longer prediction horizons,
but also to demonstrate the robustness of our approach
(which is not sensitive to minor variations of the various parameters).
 
\begin{table}%[tb]
\label{tab:EventDuration}
\caption{This table summarizes key results obtained throughout the paper.
We list all information in this one table to make it easier to make connections between 
the various quantities listed.
Description of each column.
\emph{First column}: the model or paleomagnetic reconstruction considered.
\emph{Second column}: 
number of low-dipole events in the verification portion of 
a simulation\slash paleomagnetic reconstruction.
Values in brackets are the number of events in the training data.
\emph{Third column}: maximum MCC (prediction skill) achieved for optimal WT
(see Section~\ref{sec:PredictionSkills} for the definition of MCC). 
Values in brackets are for training data.
Verification and training data are explained in Section~\ref{sec:RobustnessDurationOfTraining}
for the models and in Section~\ref{sec:SkillsData} for the paleomagnetic reconstructions.
\emph{Fourth column}: optimal WT that maximizes MCC over the training data 
(see Section~\ref{sec:ThresholdOptimization}).
\emph{Fifth column}: average duration of a low-dipole event (AED).
Values in brackets are standard deviations.
\emph{Sixth column}: average decay time (ADT) with standard deviations in brackets.
See sections~\ref{sec:QuantitativeComp} (models)
and~\ref{sec:EventDurationPaleo} (paleomagnetic reconstructions)
for definitions of average event duration and decay time and their computation.
\emph{Seventh column}: ratio of average decay time to average event duration.
All results listed here correspond to a prediction horizon $\text{PH}=1$,
a start-of-event threshold $\text{ST}=10\%$,
and an end-of-event threshold $\text{ET}=80\%$.
}
\begin{center}
\begin{tabular}{rcccccc}
&   \# of events & MCC & $\hat{\text{WT}}$ & Event duration (AED) & Decay time (ADT)& $\rho = \frac{\text{ADT}}{\text{AED}}$ \\\hline
G12 & 554 (5) & 0.96 (0.97) & 30.75\% & 3.2 kyr (0.1 kyr) & 26.9 kyr (2.0 kyr) & 8.49		\\
P09	& 551 (5) & 0.57 (0.56) & 54.50\% & 6.0 kyr (3.9 kyr) & 10.8 kyr (5.0 kyr) & 1.81        \\
DW	& 551 (5) & 0.31 (0.30) & 69.25\% & 16.1 kyr (12.2 kyr) & 8.5 kyr (4.7 kyr) & 0.53	\\
3D	& 368 (5) & 0.12 (0.14) & 17.50\% & 16.4 kyr (11.6 kyr) & 5.7 kyr (4.2 kyr) & 0.35	\\\hline
PADM2M	& 2 (4) & 0.62 (0.73) & 50.75\% & 11.7 kyr (8.1 kyr) & 25.5 kyr (10.9 kyr) & 2.19\\        
Sint-2000	& 2 (4) & 0.44 (0.77) & 36.75 \% & 10.2 kyr (8.7 kyr) & 32.0 kyr (15.1 kyr) & 3.15	
\end{tabular}
\end{center}
\label{tab:MainTable}
\end{table}%

\subsection{Finding thresholds via maximization of skill scores}
\label{sec:ThresholdOptimization}
For a fixed prediction horizon,
we compute an optimal warning threshold as follows.
For a given dipole time series,
we compute a skill score for varying WTs,
using a regular grid with spacing of $0.25\%$.
The WT that leads to the largest skill score
is selected as the optimal WT: $\hat{\text{WT}} = \text{arg max } \text{Skill} (\text{WT})$.

This approach can be implemented with a variety of skill scores,
e.g., MCC, CSI or $\F$.
For the short prediction horizons we consider,
we did not notice any significant differences 
in the optimal WTs one finds regardless of which skill score is used,
with the exception of the ACC score, which is not robust with respect to imbalances in the data
(one event occurring more often than the other).
Below we present results obtained by using MCC,
because it recently has been reported to be more appropriate than the $\F$ score 
for binary classification \citep{CJ20},
but other skill scores (not ACC) may be used to obtain similar results.

To prevent overfitting, it is necessary to validate a prediction strategy by 
applying it to an independent data set.
An optimal WT is determined by using a given dipole time series,
which we call the training data set.
The optimal WT is then applied to an independent time series,
which we call the verification data set,
and the skill score is computed for the verification data.
For the simplified models (G12, P09, DW),
the verification data are independent simulations
(using different initial conditions in the case of the deterministic G12 model
and different initial conditions and random forcing in the case of the stochastic P09 and DW models).
For the 3D model, we compute the optimal WT
by using only a portion of the simulation as training data,
and then use the remainder of the simulation as verification data.

\section{Application to a hierarchy of models}
We apply threshold-based predictions to the models in the hierarchy.
For each model, we predict a low-dipole event about as far ahead of time 
as one expects the event will last.
In our terminology, this means that the prediction horizon is equal to one average event duration
(PH $= 1 \times$ AED), but we also consider slightly longer ($1.5\times$) 
and slightly shorter ($0.5\times$) PHs. 
We also test if useful threshold-based predictions can be made
if the training period is short and, therefore, contains only a small number of low-dipole events.

\subsection{Skill of threshold-based predictions}
\label{sec:Results_Skill}
\subsubsection{Qualitative comparison and illustration}
\begin{figure}%[tb]
	\centering
	\includegraphics[width=1\textwidth]{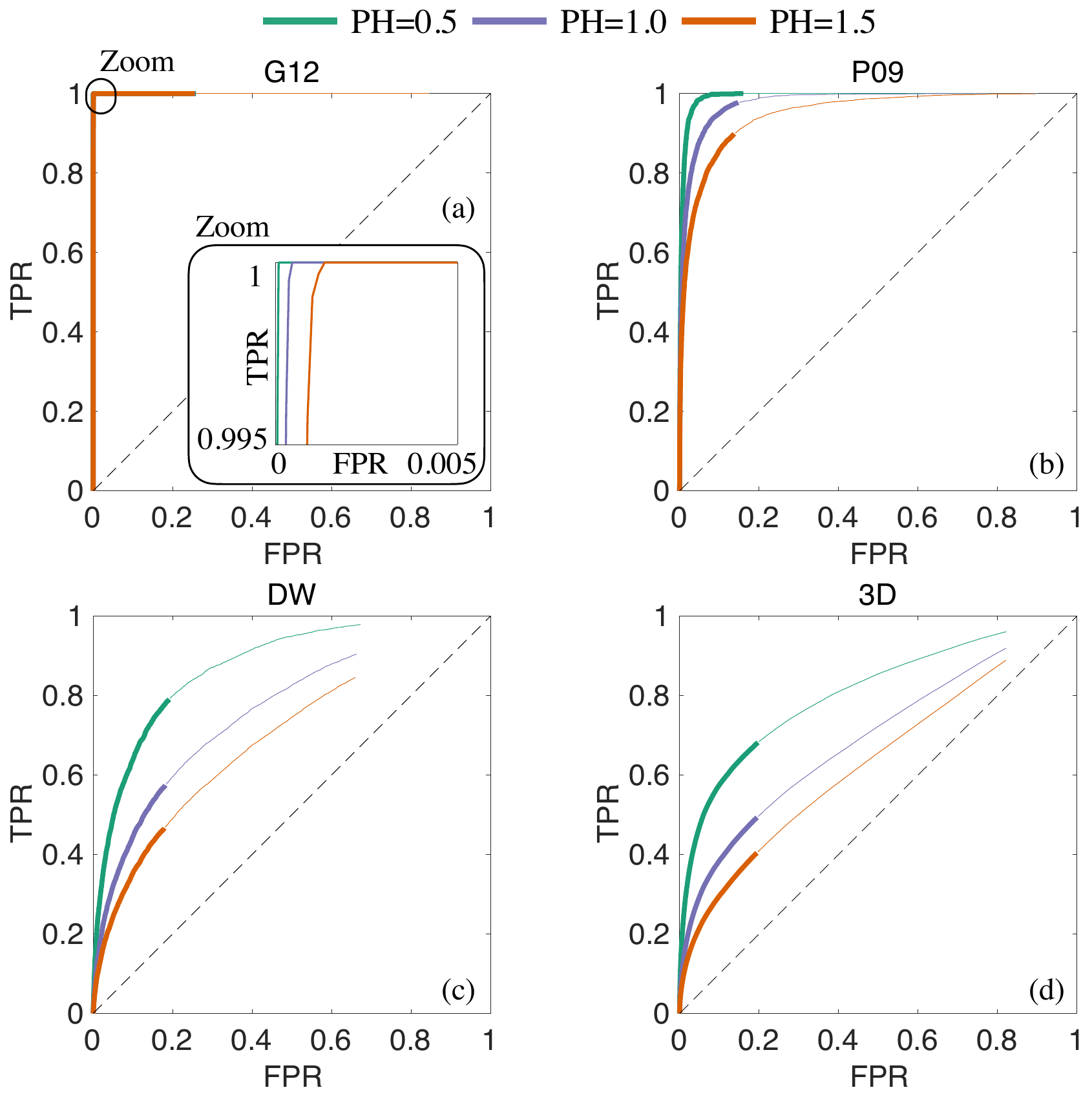}
	\caption{ROC curves for the four models and three prediction horizons,
	with $\text{PH}=0.5$ in green, $\text{PH}=1$ in purple,
	and $\text{PH}=1.5$ in orange.
	(a) G12, (b) P09, (c) DW, (d) 3D.
	A ROC curve is the collection of TPR\slash FPR pairs
	one obtains when varying the warning threshold.
	The thicker line corresponds to TPR\slash FPR pairs
	for which $\text{ST}<\text{WT}<\text{ET}$.
	The thin lines continue the ROC curves for $\text{WT}\geq \text{ET}$.
	The figure-in-figure in (a) (ROC curves for G12)
	shows a zoom near the (0,1) point to illustrate that the three ROC curves,
	corresponding to different PHs, do not overlap.
	The ROC curves are computed using long simulations,
	containing a large number of low-dipole events (see text for details).}
	\label{fig:ROC_PH}
\end{figure}

We first qualitatively assess threshold-based predictions by inspection of ROC curves.
The ROC curves shown in Figure~\ref{fig:ROC_PH}
are computed using the entire model run in the case of the 3D model,
and long simulations with around 550 events for the G12, P09 and DW models,
see Section~\ref{sec:ScalingTimeAndThresholds}.
We note that, for all four models,
the ROC curves get closer to the chance line 
(higher false positive rate, lower true positive rate)
as the prediction horizon increases.
This implies that the predictions get worse, by any measure,
as the prediction horizon increases.
This means, perhaps not so surprisingly,
that predictions via a threshold-based strategy are
more difficult to do when the prediction horizon is large.
More interestingly, we note that for any fixed PH,
the ROC curves of the G12 or P09 models
are further from the chance line than the ROC curves of the DW and 3D models.
This suggests a ``ranking'' of the models in terms of how skillful threshold-based predictions are
\footnote{But a higher ranking in predictive skill does not imply that the model is ``better,''
i.e., more similar to Earth's axial dipole, see Section~\ref{sec:Application}.}.
We investigate this ranking quantitatively via MCC skill scores below.

\begin{figure}%[tb]
	\centering
	\includegraphics[width=1\textwidth]{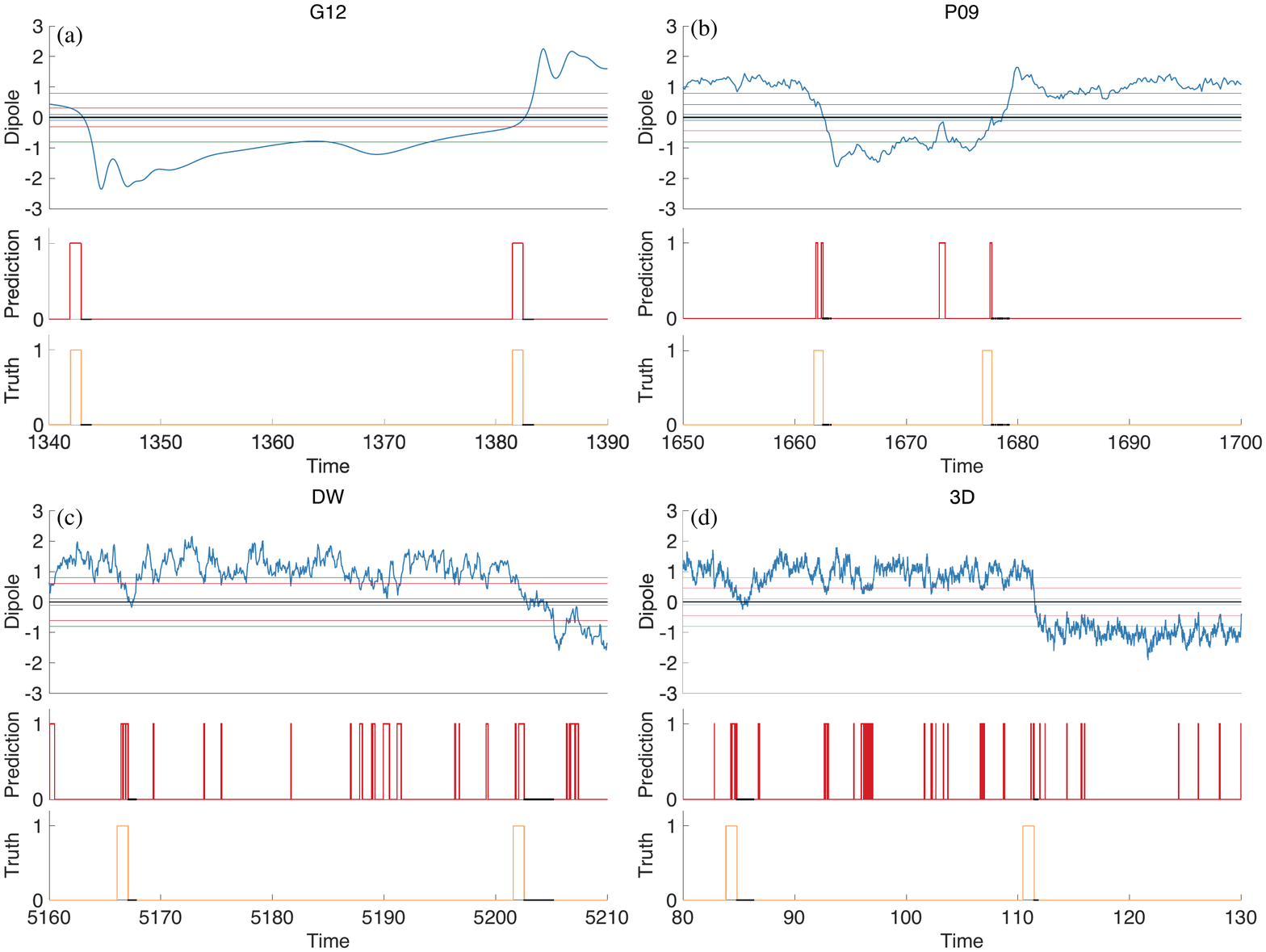}
	\caption{Illustration of threshold-based predictions for the four models.
	(a) G12, (b) P09, (c) DW, (d) 3D.
	The plots show predictions over a time period of 50 dimensionless time units
	and for a prediction horizon of $\text{PH}=1$.
	The corresponding (optimal) warning thresholds 
	(expressed in percent of the average intensity) are
	$\hat{\text{WT}}_\text{G12} = 31.25\%$, 
	$\hat{\text{WT}}_\text{P09} = 43.00\%$, 
	$\hat{\text{WT}}_\text{DW} = 60.25\%$ and 
	$\hat{\text{WT}}_\text{3D} = 45.50\%$ 
	for respectively the G12, P09, DW and 3D models.
	These optimal WTs are computed using time series
	of the four models that contain a large number of events (see text for details).
	For each model, a dimensionless time is defined via a scaling of time with the average 
	event duration (see text for details).
	Each sub-figure contains three panels.
	\emph{Top}.
	Blue: dipole time series.
	Blue\slash green\slash red horizontal lines: 
	start-of-event\slash end-of-event\slash warning thresholds.
	\emph{Center}.
	Graphs are zero if the threshold-based prediction
	is ``no low-dipole event will start during the prediction horizon;''
	graphs are one if the threshold-based prediction
	is ``a low-dipole event will start during the prediction horizon.''
	\emph{Bottom}.
	The graphs are zero if no low-dipole event starts
	within the prediction horizon;
	the graphs are one if a low-dipole event starts during the prediction horizon.
	Black lines in the center and bottom graphs denote times when no predictions are being made.
	}
	\label{fig:Prediction_Illu}
\end{figure}

For each model, we illustrate threshold-based predictions for which an optimal
warning threshold is found by maximizing MCC skill score,
as a function of the warning threshold.
The data sets used for finding the optimal WTs
are the entire model run in the case of the 3D model,
and long simulations with around 550 events for the G12, P09 and DW models,
see Section~\ref{sec:ScalingTimeAndThresholds}
(no distinction between training and verification data).
This results in optimal WTs of $\hat{\text{WT}}_\text{G12} = 31.25\%$, $\hat{\text{WT}}_\text{P09} = 43.00\%$,
$\hat{\text{WT}}_\text{DW} = 60.25\%$ and $\hat{\text{WT}}_\text{3D} = 45.50\%$
for respectively the G12, P09, DW and 3D models.
Results for a prediction horizon $\text{PH} =1$
are shown in Figure~\ref{fig:Prediction_Illu};
results for $\text{PH} =0.5$ or $\text{PH} =1.5$ are similar.
We plot excerpts of the dipole time series of the four models,
along with two graphs that illustrate the predictions and their validity.
Each model is represented by one sub-figure which contains three panels.
The top panel shows an excerpt of the dipole time series.
We show a time interval of 50 non-dimensional time units for each model
and each model exhibits two events during this time interval
(recall that time is scaled by the average event duration, AED, see Table~\ref{tab:MainTable}).
The orange lines in the bottom of each sub-figure
are zero if no low-dipole event starts within the prediction horizon,
and they are one if a low-dipole event starts during the prediction horizon.
Because we show the same time-interval in non-dimensional units,
the intervals during which the orange line is at one are of equal width across all four sub-figures.
The red lines in the center panels are
zero if no low-dipole event is predicted to start during the prediction horizon;
the lines are one if a low-dipole event is predicted to start during the prediction horizon.
Thus, the overlap of the red and orange lines defines TPs, FPs, TNs and FNs,
and a large overlap corresponds to a skillful prediction.
For example, a FP corresponds to a situation where the orange is at zero while the red line is at one;
a FN corresponds to a situation where the orange line is at one while the red line is at zero.

We note that predictions for the G12 model
lead to a small number of false positives or false negatives.
In the excerpt shown for the G12 model in Figure~\ref{fig:Prediction_Illu},
there is only one false positive, caused by the prediction starting one time step too early
(during the first of the two events shown).
For the DW and 3D models, on the other hand,
we note a large number of false positives and false negatives,
which renders threshold-based predictions unreliable for these models.
Comparisons of the graphs for G12 and the DW and 3D models
suggest that threshold-based predictions for the DW or 3D model
are indeed worse, by any measure, than those for the G12 model.
In the case of P09,
we note a larger number of false positives and false negatives
than in the case of G12,
but false positives or false negatives occur less frequently than for the DW or 3D models.
Consistent with what was suggested by Figure~\ref{fig:ROC_PH},
the skill of threshold-based predictions for the P09 model thus 
seems to fall in between the skills of threshold-based predictions 
for the G12 (very high skill) and DW\slash 3D models (very low skills).

\subsubsection{Quantitative comparison and ranking}
\label{sec:QuantitativeComp}
We compute MCC skill scores to quantitatively compare the skill
of threshold-based predictions for the various models.
To avoid over-fitting we now compute skill scores on verification data,
i.e., data that are \emph{not} used for computing the optimal warning threshold,
as described in Section~\ref{sec:ThresholdOptimization}.

We generate training and verification data as follows.
For the G12, P09 and DW models,
the training data are the long simulations that were also used
in Section~\ref{sec:ScalingTimeAndThresholds}.
The verification data are ten independent simulations, each of length  $10^4$.
For the 3D model, we create training and verification data
by ``chopping up'' the overall simulation as follows.
We split the simulations into two parts of equal length
and use one for training and the other for verification.
We then repeat the procedure, 
but split the simulation into three equally long portions,
using one for training and two for verification.
Finally, we split the simulation into four equally long portions,
and use one for training and three for verification.
This procedure leads to six MCC scores over verification data.
Generating multiple verification data sets in this way allows us to estimate the variability  
in the skill of threshold-based predictions for all four models.

\begin{figure}%[tb]
	\centering
	\includegraphics[width=0.5\textwidth]{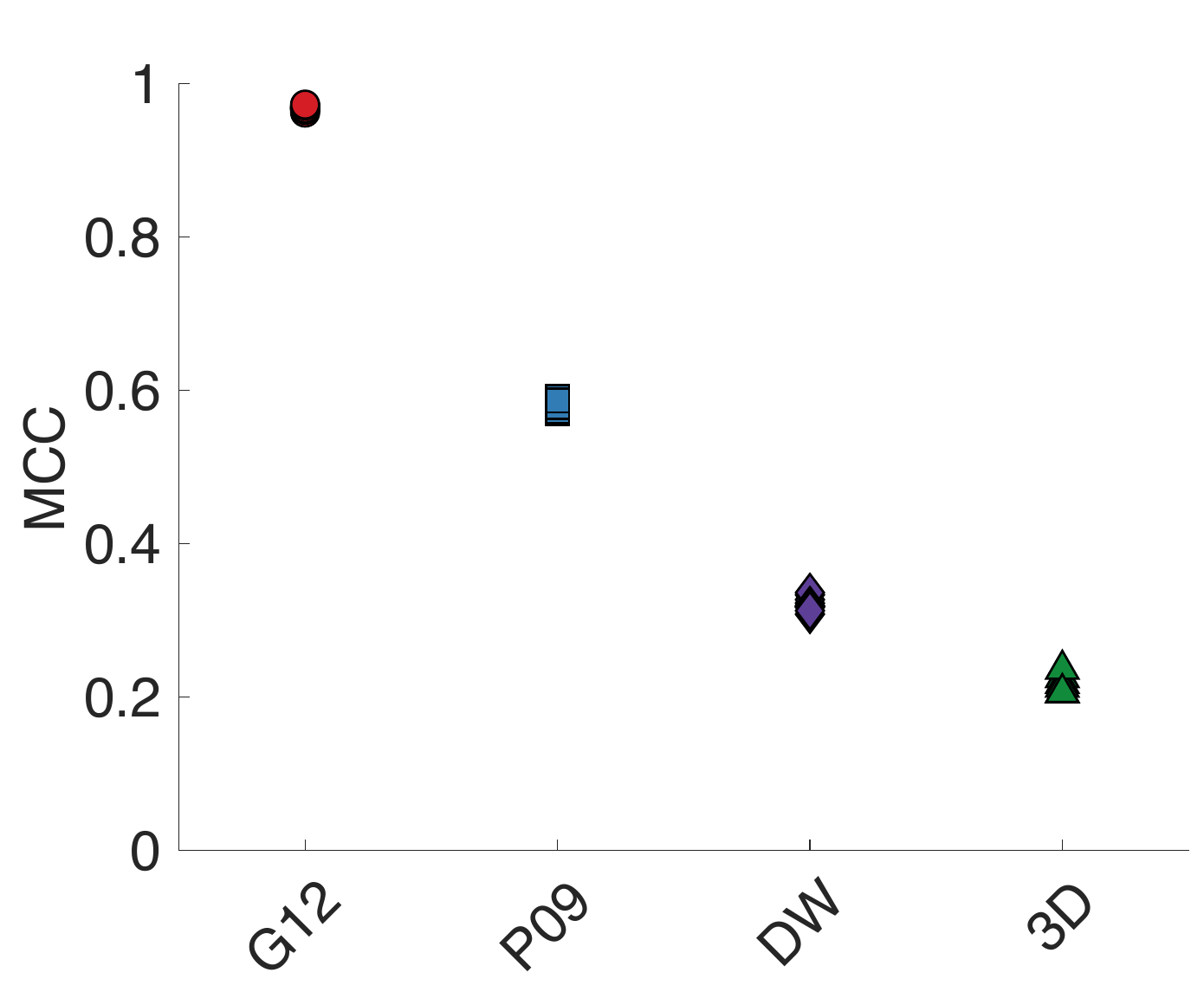}
	\caption{MCC skill scores (verification) of the four models
	for prediction horizon $\text{PH}=1$.
	For each model, several MCC scores are shown.
	The MCC scores are computed over multiple sets
	of verification data. The training data are long simulations containing many low-dipole events (see text for details).}
	\label{fig:Skill_PH}
\end{figure}

Results are shown in Figure~\ref{fig:Skill_PH},
where we plot MCC scores for the four models for threshold-based
predictions with prediction horizon $\text{PH=1}$.
We only show the results for a prediction horizon $\text{PH}=1$,
but one obtains qualitatively the same results with $\text{PH}=0.5$ or $\text{PH}=1.5$.
We note that the variation in skill over the different verification data sets is small.
This suggests that the verification data sets are ``large enough'' 
so that variation in the verification data does not affect the scores.
Moreover, our results confirm the ranking of the skill of threshold-based predictions
that we anticipated from inspection of ROC curves.
Specifically, we rank the models (skill from high to low)
in terms of their predictability via intensity thresholding as:
 G12, P09, DW and 3D.
Indeed, we found that this result is independent of the choice of skill score --
one obtains qualitatively and, to a large extent, quantitatively
the same results using, e.g., the $\F$ or CSI skill scores.

\begin{figure}%[tb]
	\centering
	\includegraphics[width=1\textwidth]{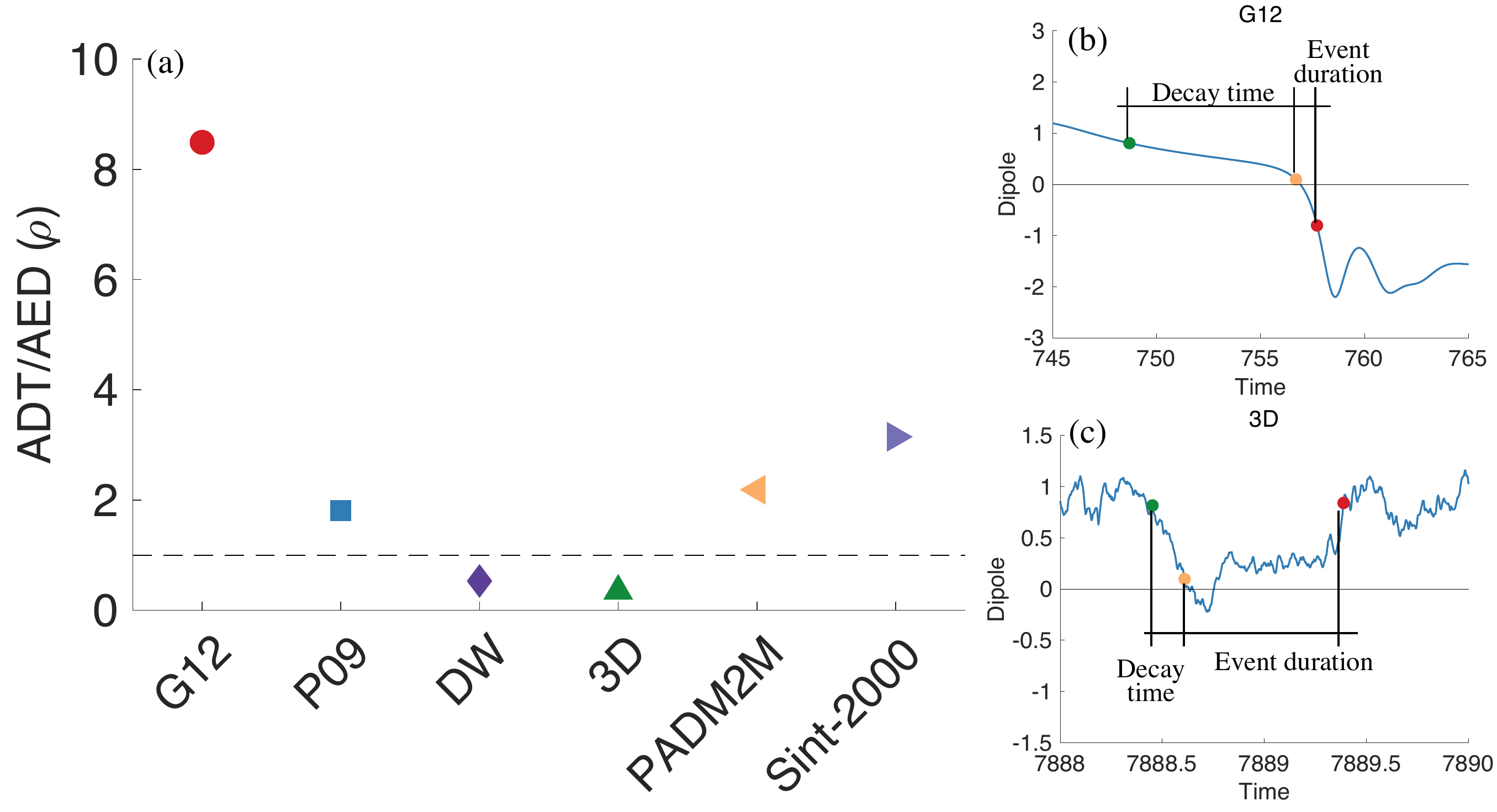}
	\caption{
	(a): ratio $\rho$ of the average decay time (ADT) to the average event duration (AED) for the four models
	and two paleomagnetic reconstructions (PADM2M and Sint-2000, see Section~\ref{sec:Application}).
	Also shown is the $\rho=1$ line (dashed).
	(b): illustration of the decay time and event duration of an event for G12.
	(c): illustration of the decay time and event duration of an event for the 3D model. 
	The beginning of the decay is marked in green, the start of an event is marked in orange and the end of an event is marked in red. 
	The decay time is the time interval between the start of the decay and the start of an event. 
	The event duration is the time interval between the start and end of an event.
	}
	\label{fig:DecayTime}
\end{figure}

This ranking and, more generally, 
the skill of threshold-based predictions 
appears to be determined by an interplay of:
\begin{enumerate}[(i)]
\item
\emph{The extent of variation in the dipole intensity}:
a high potential for false positives results if 
the intensity dips to low values regularly, but if no low-dipole event follows.
\item
\emph{The decay rate prior to a low-dipole event}:
a quick decay results in a high potential for false negatives.
\end{enumerate}
For (i), 
we recall the intensity histograms of Figure~\ref{fig:AmpHists},
which show that the P09, DW and 3D models (low skill) spend more time
at low intensity values than the G12 model (high skill).
For (ii), 
we compute the \emph{average decay time} (ADT),
which measures how quickly the dipole intensity decays
prior to a low-dipole event.
We define the \emph{decay time} as the 
absolute value of the time difference between
the start of the event and the last previous instance at which the field exceeded the end-of-event threshold (ET, 80\% of its average value).
We list the ADT for the four models in Table~\ref{tab:MainTable}, with standard deviations.
These ADT should be compared to the average event durations (AED), also listed in Table~\ref{tab:MainTable}.
Recall that the event duration is 
defined by the time interval
that starts when the dipole intensity drops below a given start-of-event threshold
(10\% of the average value)
and ends when the dipole intensity exceeds a given end-of-event threshold
(80\% of the average value).
Thus, the average event duration
describes how quickly on average the dipole recovers to a large value
after it dropped to a low value.
If the average decay time is larger than the average event duration (ADT $>$ AED),
then low-dipole events occur slowly;
if the average decay time is smaller than the average event duration (ADT $<$ AED),
then low-dipole events occur quickly.
The two behaviors are illustrated by the G12 and 3D models
in Figures~\ref{fig:DecayTime}(b) and~\ref{fig:DecayTime}(c).

In Figure~\ref{fig:DecayTime}(a),
we plot the ratio of the average decay time
to the average event duration 
for the four models (see Table~\ref{tab:MainTable}).
For brevity, we introduce an abbreviation for this ratio:
\begin{linenomath}\begin{equation}
	\rho = \frac{\text{ADT}}{\text{AED}}.
\end{equation}\end{linenomath}
We note that $\rho$
follows a similar trend as the MCC score.
In particular, the ranking of the models it leads to is identical to the ranking inferred from the MCC skill score.
This suggests that the skill of threshold-based predictions is 
influenced by how quickly low-dipole events occur with respect to their duration.
If they occur slowly (ADT $>$ AED, $\rho>1$),
then threshold-based predictions have a high skill.
If they occur quickly (ADT $<$ AED, $\rho<1$),
then  threshold-based predictions may have a low skill.

\subsection{Robustness of skill to a short training period}
\label{sec:RobustnessDurationOfTraining}
\begin{figure}%[tb]
	\centering
	\includegraphics[width=1\textwidth]{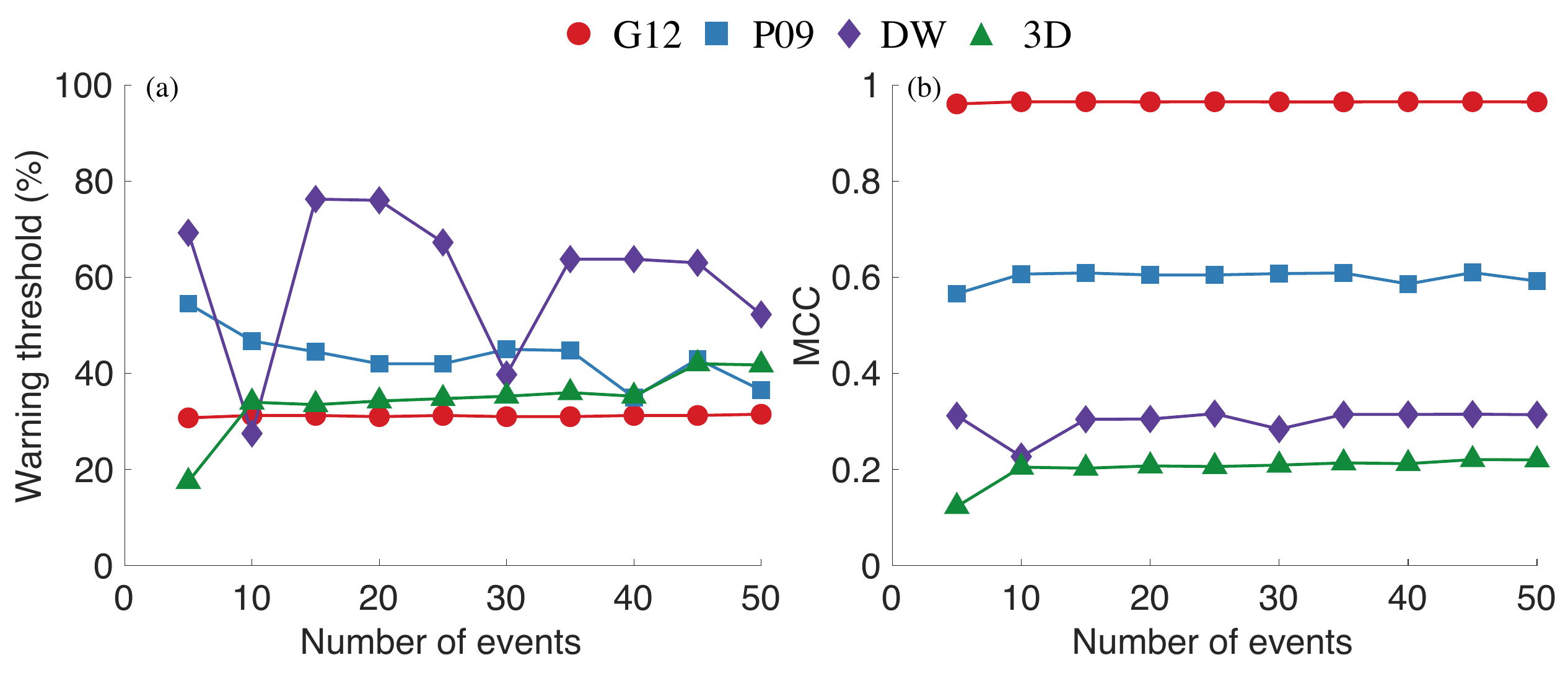}
	\caption{(a) optimal warning threshold as a function of the number of events contained in the training data. 
	(b) MCC computed over verification data as a function of the number of events contained in the training data. 
	The prediction horizon is $\text{PH}=1$.
	}
	\label{fig:Robustness_Training}
\end{figure}

Motivated by the fact that the observational record is short 
(the PADM2M and Sint-2000 reconstructions that we investigate in Section~\ref{sec:Application}
extend over 2 Myr and contain only six low-dipole events),
we investigate the robustness of the optimal WT and corresponding skill
with respect to the duration of the training data.
For each model we compute
an optimal WT for several training data sets of different durations,
and, hence, containing a different number of events.
Results for a prediction horizon $\text{PH}=1$ are shown in Figure~\ref{fig:Robustness_Training}(a).
For G12, we note that the optimal WT is nearly independent of the duration of the training data set.
This means that, for this model, one can find a useful warning threshold
from a rather short training period.
For all other models, we observe a variation of the 
optimal warning threshold as we vary the duration of the training period.
The variations are most significant for the DW model, 
for which the optimal WT varies from about 25\% to nearly 80\% (which is the maximum allowed value).
For P09, the optimal WT varies between 35\% and 55\%,
but there seems to be a plateau of nearly constant WT for training data sets that contain 20-35 events.
For the 3D model, we observe a variation of the optimal WT between 20\% to about 40\%.
Again, we note a plateau of nearly constant WT for training data with 10-40 events.

Variations in the optimal WT, however, do not necessarily imply variations
in the resulting MCC skill score.
This is shown in Figure~\ref{fig:Robustness_Training}(b).
Here, we use the optimal WTs obtained from the same various training periods (and shown in
Figure~\ref{fig:Robustness_Training}(a)),
but compute the MCC over the verification data.
For the simple models (G12, P09 and DW),
the verification data are the long simulations (with about 550 events, see Section~\ref{sec:ScalingTimeAndThresholds}).
For the 3D model,
the verification data are the portion of the simulation that was not used during training.

We observe that the MCC skill score of threshold-based predictions
is nearly independent of the duration of the training data.
This is consistent across the hierarchy of models
and suggests that the shortness of the observational record
may not be the critical limiting factor for determining a useful warning threshold.
Our numerical results indeed suggest that a useful WT can be found
even if the training period is short and comparable with the observational record.

\begin{figure}%[tb]
	\centering
	\includegraphics[width=1\textwidth]{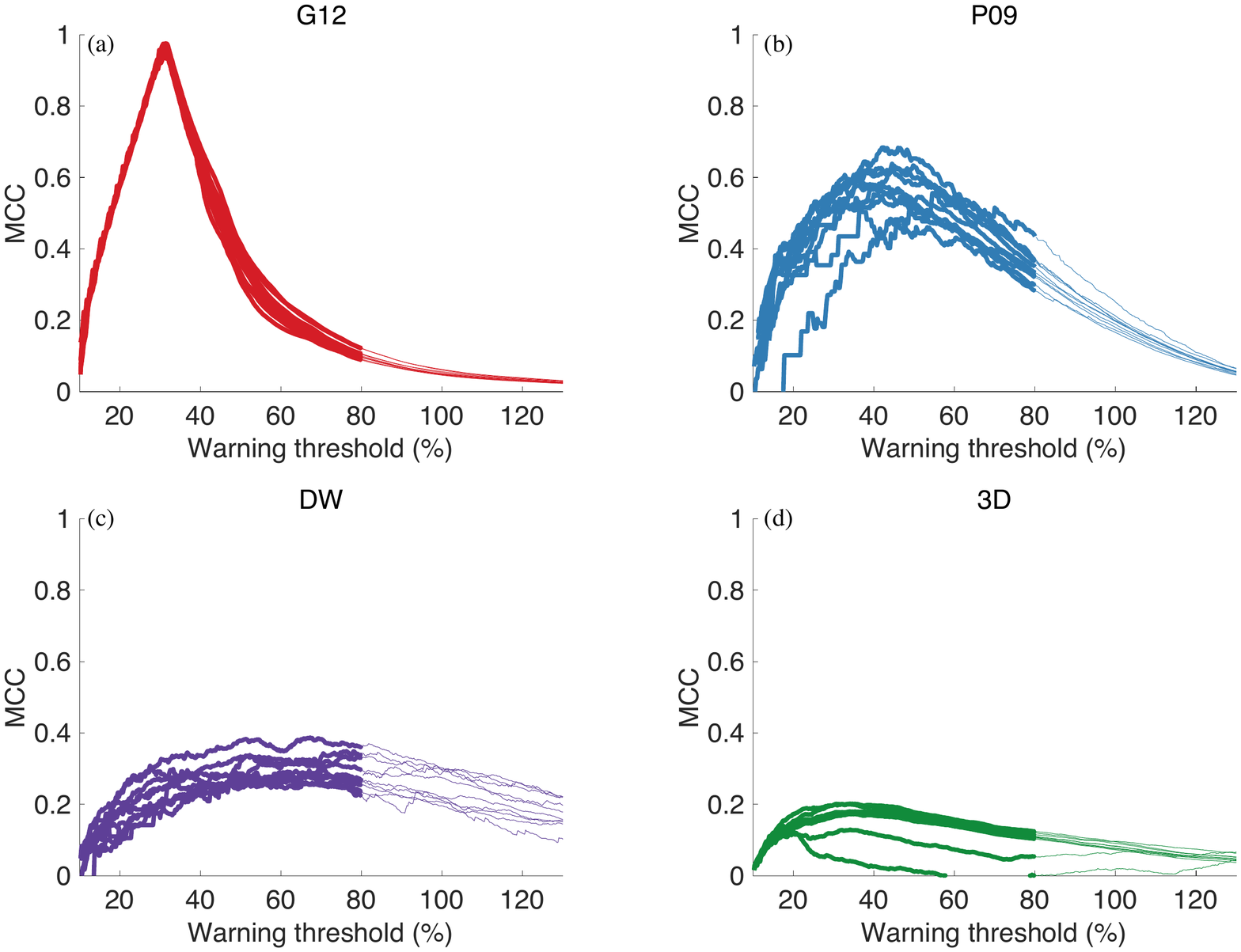}
	\caption{
	MCC skill score as a function of WT for the four models.
	(a) G12, (b) P09, (c) DW, (d) 3D.
	The various graphs shown for each model differ
	in the number of events contained in the training data
	(see text for details). The thin lines continue the curves for $\text{WT}\geq \text{ET}$.}
	\label{fig:Robustness_Training_MCC}
\end{figure}

The reason \emph{why} the skill is independent of the duration of the training data varies across the hierarchy.
This can be understood by considering how MCC depends on WT,
which we compute and show in Figure~\ref{fig:Robustness_Training_MCC}.
If the MCC vs.~WT graph is sharply peaked around an optimal value,
and if the peak is nearly independent of the duration of the training period,
then a good WT can be found even with a limited amount of training data.
This is the case for the G12 model.
If the graph of MCC skill score plateaus for large values of WT,
then rather different WT values can produce a similar skill scores.
This is the explanation for why drastic variations in optimal WT
cause nearly no variations in the resulting optimal MCC
in case of the DW model in Figure~\ref{fig:Robustness_Training}.

%-----------
\subsection{Impact of data filtering}
Threshold-based predictions for the 3D and DW model
have a low skill compared to P09 or G12.
This could be due to the quick changes in polarity
that we observe in the 3D and DW models, and that occur on short time scales (recall Figure~\ref{fig:Models}).
These are absent from the P09 or G12 models.
Palomagnetic reconstructions such as PADM2M and Sint-2000,
which are inherently smoothing the field they record through the slowly depositing sedimentary process,
also fail to show such a behavior.
One may thus wonder if the 3D or DW models could
become more amenable to threshold-based predictions if the dipole is smoothed in an analogous way.

To test this possibility, we first consider the 3D model, and rely on the secular variation time scale $\tau=415$ years,
which we already used to scale time for this simulation.
The idea is to test a filtering that mimics the sedimentary process and makes physical sense
from the point of view of a 3D dynamo. For 3D dynamos, and for Earth's dynamo, the secular variation time scale
defines the main time scale with which the non-dipole field is behaving \citep{lhuillier2011grl}.
It provides a natural separation between the times scales of the long-term behavior of the dipole field,
which is the one we are most interested in here, and its short time scales.
Smoothing over a time period of $4\tau$ typically removes such short time scales \citep{Hulot:1994}.
This corresponds to about 2 kyr.
This is the value we tested, as it also is roughly consistent with the smoothing due to the sedimentary process in paleomagnetic reconstructions
such as PADM2M and Sint-2000.
For example, the regularization used to obtain the PADM2M reconstruction
suppresses energy at time scales of 5-10 kyr \citep{PADM2M}.
It finally is short enough compared to the decay time and event durations we identified for the field
produced by the 3D (and DW) model (see Table~\ref{tab:MainTable}).
For consistency, we then also used the same time filtering to filter the time series produced by the DW model.
In both cases, we used a moving average filter.
Results are provided in Table~\ref{tab:SmoothingTable},
which lists the optimal MCC of threshold-based predictions for the 
DW and 3D models with and without smoothing for three prediction horizons.
We found that the
skill of threshold-based predictions only slightly increases for the 3D model, 
but hardly at all (to two digits) for the DW model.
Thus, skills associated with the DW and 3D models are nearly unchanged by the smoothing process,
and remain smaller than the skills associated with the P09 and G12 models.

\begin{table}%[tb]
\caption{Maximum MCC of threshold-based predictions for the DW and 3D models with and without smoothing (smoothing window is $4\tau\approx2\text{kyr}$) 
for three different prediction horizons.
Optimal warning thresholds and MCC scores
are computed over the entire run (no verification).}
\begin{center}
\begin{tabular}{rrccc}
&Prediction horizon	& 0.5 & 1 & 1.5 \\\hline	 
\multirow{2}{*}{DW}	& No smoothing	 & 0.39 & 0.32 & 0.27 \\
				& 2 kyr smoothing	 & 0.39 & 0.32 & 0.27\\				
\multirow{2}{*}{3D}	& No smoothing	& 0.28 & 0.22 & 0.18 \\
				& 2 kyr smoothing	& 0.32 & 0.24 & 0.19 
\end{tabular}
\end{center}
\label{tab:SmoothingTable}
\end{table}%

\subsection{Summary of results from the hierarchy of models}
The hierarchy of models is consistent in 
that threshold-based predictions become more difficult,
or, equivalently, less skillful, when the prediction horizon increases.
This suggests that threshold-based predictions are at best useful
for predicting low-dipole events with a lead time that is comparable to the average duration of the event
(about 10 kyr on Earth's time scales).
Moreover, the machinery of identifying thresholds by maximizing a skill score
is robust in the sense that the skill during training is comparable to the skill during verification.
Our overall approach is also robust with respect to the precise choices
of start-of-event and end-of-event thresholds,
and with respect to the choice of skill score (MCC, $\F$ or CSI).

We observe strong differences in the skills of threshold-based predictions across the various models.
The DW and 3D models exhibit complex behavior during reversals or excursions,
with many polarity changes during the low-dipole event
and the decay time is short compared to the event duration (fast reversals).
The G12 model behaves differently: 
we do not observe quick polarity changes during a G12 reversal,
no major excursions occur,
and  the decay time is larger than the event duration (slow reversals).
The G12 model is more amenable to threshold-based predictions than the DW or 3D models,
because of its simpler reversing behavior and because reversals are approached slowly.
The P09 model falls in between the DW and 3D models and the G12 model.

Our numerical experiments with short training data sets,
suggest that the main difficulty for threshold-based predictions
may not be the shortness of the observational record.
The hierarchy of models is surprisingly consistent
in that one may be able to determine useful warning thresholds,
even if the training data are limited.
The reasons for \emph{why} this stability occurs,
however, vary across the hierarchy of models.
For the G12 model, low-dipole events are indeed easy to predict by a threshold
and this threshold can be found by optimizing skill scores over short data sets.
For the other models, the skill score is a nearly flat function of the threshold,
i.e., different thresholds can lead to similar skill scores (recall Figure~\ref{fig:Robustness_Training_MCC}).
More importantly, the overall skill of threshold-based predictions is low for the DW and 3D models,
even when introducing some smoothing.
Thus, threshold-based predictions may be of limited use for the DW and 3D models,
because false positives and false negatives occur frequently.
Again, the P09 model falls in between the G12 and DW\slash 3D models.

We summarize our main results about threshold-based predictions for dipole models as follows.
\begin{enumerate}[(i)]
\item
Across the hierarchy of models,
the skill of threshold-based predictions degrades with the prediction horizon.
\item
Across the hierarchy of models,
threshold-based predictions are robust to minor variations of numerical details,
such as choice of skill sore (MCC or F1 or CSI), or choices of start-of-event and end-of-event thresholds.
\item 
Across the hierarchy of models,
useful warning thresholds can be found
even if the duration of the training period is short
and comparable to the observational record.
This suggests that the shortness of the observational record is not the main issue 
that makes computing warning thresholds difficult.
The reasons for \emph{why} this is the case, however, differs across the hierarchy of models.
\item 
The G12 model
is more amenable (highest skill) to threshold-based predictions than 
the DW or 3D models (lowest skill).
%We found a partial explanation for why this is the case.
The skill of threshold-based predictions for the P09 model
falls in between the skills for G12 and DW\slash 3D.
Furthermore, we found that skills strongly correlate with the ratio of the average decay time to the average event duration.  
\end{enumerate}

\section{Application to paleomagnetic reconstructions}
\label{sec:Application}
We now take advantage of the lessons learned from the hierarchy of models and apply
threshold-based predictions to the PADM2M and Sint-2000 paleomagnetic reconstructions,
which provide proxies of the Earth's axial dipole intensity over the past 2 Myr \citep{PADM2M,SINT2000}.
More specifically,
PADM2M and Sint-2000 report the virtual axial dipole moment (VADM)
in increments of 1 kyr for the past 2 Myr.
We scale each reconstruction so that one unit of relative paleointensity
corresponds to its time average 
($5.32\cdot 10^{22}$ Am$^2$ for PADM2M, $5.81\cdot 10^{22}$ Am$^2$ for Sint-2000).
The timing of reversals is based on the geomagnetic polarity time scale of \cite{CK95},
with a slight modification for the Cobb mountain sub-chron in the case of PADM2M \citep{MFH17}. 

We note that PADM2M and Sint-2000 are ``data'' of the same process,
namely Earth's dipole intensity over the past 2 Myr.
Nonetheless, there are differences between PADM2M and Sint-2000,
which are due to variations in the processing and interpretation of raw data,
and also the raw data that goes into the two reconstructions.
This means that differences between PADM2M and Sint-2000 
indicate the level of uncertainty that is caused 
by difficulties with observing Earth's dipole over millions of years
(see also \cite{MB19}). 
Moreover, the fact that the observational record is short (2 Myr sampled in 1 kyr increments),
implies that it is difficult to determine if any differences are (statistically) significant.
It is important to keep this ``minimum level of uncertainty'' in mind 
when evaluating threshold-based predictions for the paleomagnetic reconstructions
(note that we essentially treat the paleomagnetic reconstructions as ``data,''
but we are aware that these reconstructions are themselves ``models''). 

\subsection{Event durations and decay times}
\label{sec:EventDurationPaleo}
Based on our definitions above,
we compute the average and standard deviation of the event duration 
and decay times for
the six events of PADM2M and Sint-2000.
Results using ST = 10\% and ET = 80\% as before,
are listed in Table~\ref{tab:MainTable}
and these values should be compared with the corresponding values for the four models.
In this context, it is important to realize that 
PADM2M and Sint-2000 never exhibit intensity values below 10\% of their time average,
which is why the definition of the low-dipole event in Section~\ref{sec:Definitions}
contains the ``or-statement'':
a low-dipole event starts when the intensity drops below
the ST \emph{or} if the dipole changes its sign.

\begin{figure}%[tb]
	\centering
	\includegraphics[width=0.75\textwidth]{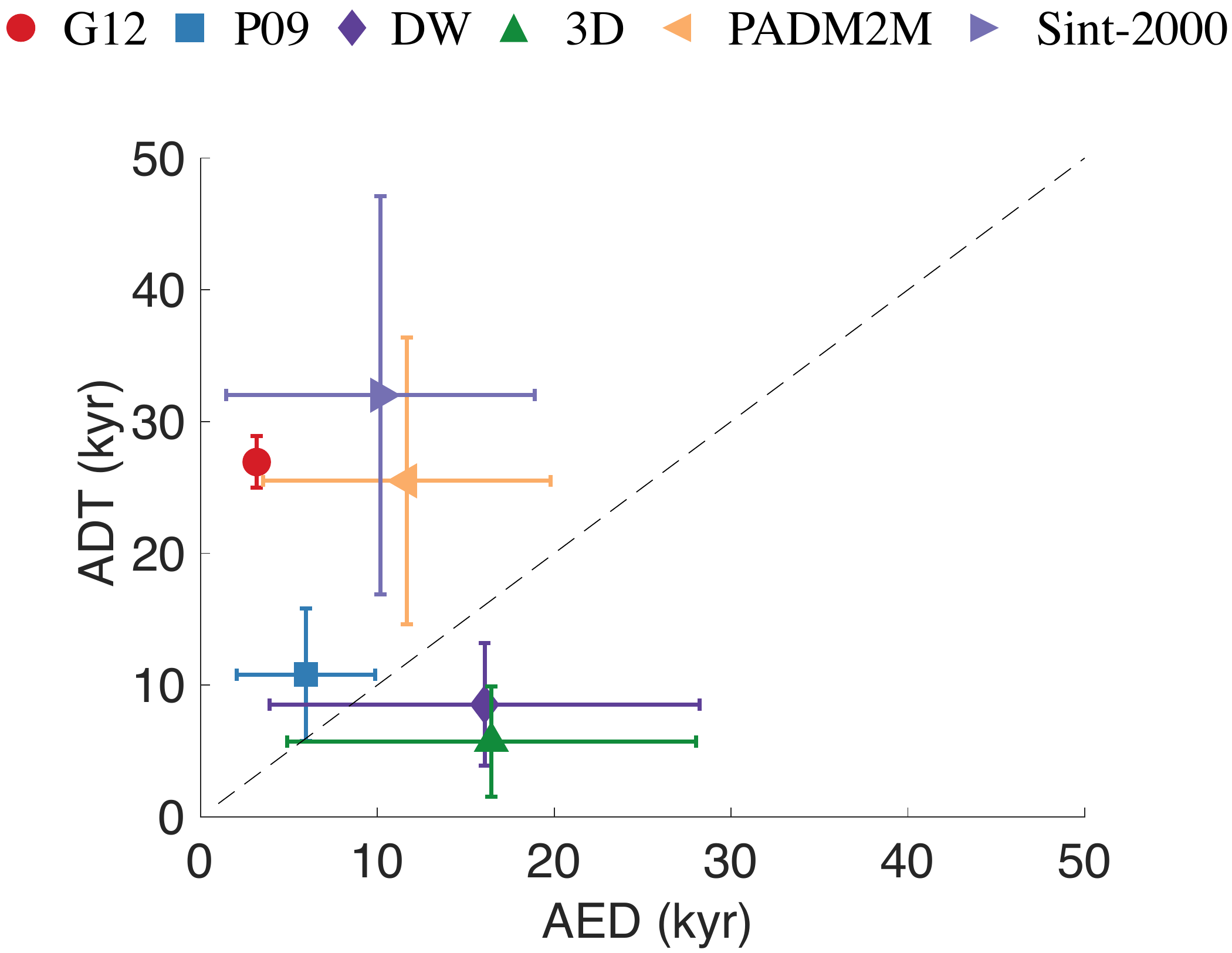}
	\caption{
	Average decay time (ADT) plotted as a function of the average event duration (AED) for the four models
	and the paleomagnetic reconstructions.
	Also shown are the error bars based on one standard deviation.
	In the case of the G12 model, the standard deviation of the event duration is too small
	to be visible as an error bar.
	Also shown is a $45^\circ$ line that separates models or data for
	which ADT $>$ AED from models for which ADT $<$ AED.
	}
\label{fig:DecayVsDuration}
\end{figure}

We first note that PADM2M and Sint-2000 lead to results consistent with each other
(e.g., average event duration and average decay times agree with each other
within the corresponding standard deviations).
We also note that both average event durations and average decay times fall within the range
of values covered by the hierarchy of models.
Hardly any model, however, leads to values satisfyingly matching those of PADM2M and Sint-2000 for both quantities.
This is best seen in Figure~\ref{fig:DecayVsDuration}, which shows the average decay time (ADT) plotted
as a function of the average event duration (AED) for the four models and the paleomagnetic reconstructions.

The average event duration of PADM2M or Sint-2000 is longer than that of the G12 (shortest) and P09 models,
and shorter than that of the DW and 3D (longest) models. 
We note, however, that associated standard deviations may reconcile
the average event durations of the paleomagnetic reconstructions with those of the various models, 
but only marginally so for G12,
which intrinsically displays little variation in the event duration.
Moreover, the standard deviations for the event durations of the paleomagnetic reconstructions 
are quite comparable 
to those of the DW and 3D models,
but larger than those of the P09 model,
and are much larger than those of the G12 model.
Overall, the average event duration
of the paleomagnetic reconstructions lies in-between the average event durations of the G12\slash P09 and DW\slash 3D models.
We keep in mind that standard deviations for the paleomagnetic reconstructions
may be corrupted by insufficient statistics, since the data document only six events.

The situation, however, is different when considering average decay times.
The average decay times of the paleomagnetic reconstructions 
are much larger than those of the 3D (shortest), DW and P09 models,
but comparable to that of the G12 model (longest). 
The standard deviations are much larger than that of the G12 model,
and substantially larger than those of the P09, 3D and DW models 
(P09, DW and 3D models are comparable).
This could be due to insufficient statistics or to data uncertainties, as suggested by the disagreement
between the different values obtained with the PADM2M and Sint-2000 data sets.
From the perspective of average decay times, it thus appears that the data are consistent with the G12 model.

Finally,  we compute the ratio $\rho$ of the average decay time to the
average event duration for both paleomagnetic reconstructions.
This leads to values of about two for PADM2M and three for Sint-2000 (see Table~\ref{tab:MainTable}),
which is consistent with the already known fact that intensity tends to decrease more slowly before a reversal than
it recovers after it \citep{SINT2000}.
The ratio $\rho$ of PADM2M and Sint-2000
can also be compared to the corresponding ratios of the four models in Figure~\ref{fig:DecayTime}.
We note that the ratios of the paleomagnetic reconstructions are much larger than the corresponding ratios associated with the 3D (smallest) and DW models;
they are comparable to the corresponding ratio of the P09 model, and much smaller than the corresponding ratio of the G12 model.

\subsection{Threshold-based predictions and their skills}
\label{sec:SkillsData}
\begin{figure}%[tb]
	\centering
	\includegraphics[width=0.9\textwidth]{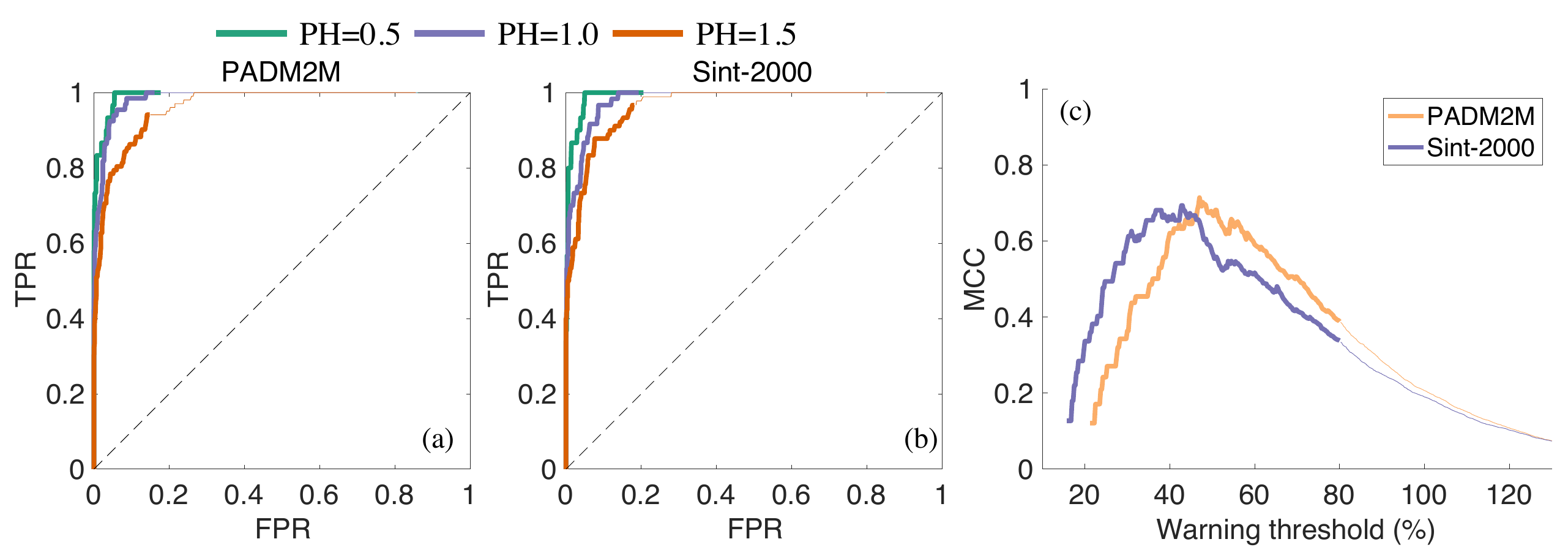}
	\caption{
	Panels (a) and (b): ROC curves for two paleomagnetic reconstructions
	and three prediction horizons,
	with $\text{PH}=0.5$ (about 6 kyr) in green, $\text{PH}=1$ (about 11 kyr) in purple,
	and $\text{PH}=1.5$ (about 17 kyr) in orange.
	(a) PADM2M, (b) Sint-2000.
	An ROC curve is the collection of TPR\slash FPR pairs
	one obtains when varying the warning threshold.
	The thicker line corresponds to TPR\slash FPR pairs
	for which $\text{ST}<\text{WT}<\text{ET}$.
	The thin lines continue the ROC curves for $\text{ET}<=\text{WT}$.
	Panel (c): MCC as a function of the warning threshold
	(prediction horizon is $\text{PH}=1$).
	The ROC curves and MCC scores are computed over the entire
	2 Myr covered by the paleomagnetic reconstructions.
	}
	\label{fig:ROC_Obs}
\end{figure}

We now apply threshold-based predictions to PADM2M and Sint-2000
using the same techniques as above and, as before, 
consider prediction horizons $\text{PH}=0.5$, $\text{PH}=1$ and $\text{PH}=1.5$.
Note that these PHs correspond to about 6 kyr, 11 kyr and 17 kyr in geophysical time.
The ROC curves of threshold-based predictions for PADM2M and Sint-2000 are shown in Figures~\ref{fig:ROC_Obs}(a) and~\ref{fig:ROC_Obs}(b).
These curves are computed over the entire 2 Myr time window covered
by the paleomagnetic reconstructions.
Inspecting the ROC curves qualitatively,
we see that the skill of threshold-based predictions decreases with the prediction horizon.
We observed this also for all four models.
Comparing the ROC curves of the paleomagnetic reconstructions in Figures~\ref{fig:ROC_Obs}(a) and (b)
with the ROC curves of the models in Figure~\ref{fig:ROC_PH},
the ROC curves of the paleomagnetic reconstructions resemble those of the P09 model.
Figures~\ref{fig:ROC_Obs}(c) shows the curve traced out by the MCC as when varying the warning threshold
for PADM2M and Sint-2000 (MCC is computed over the entire 2 Myr time window).
Again, we note that the curves corresponding to the paleomagnetic reconstructions
are qualitatively similar to the corresponding curve of the P09 model (see Figure~\ref{fig:Robustness_Training_MCC}).

\begin{figure}%[tb]
	\centering
	\includegraphics[width=.5\textwidth]{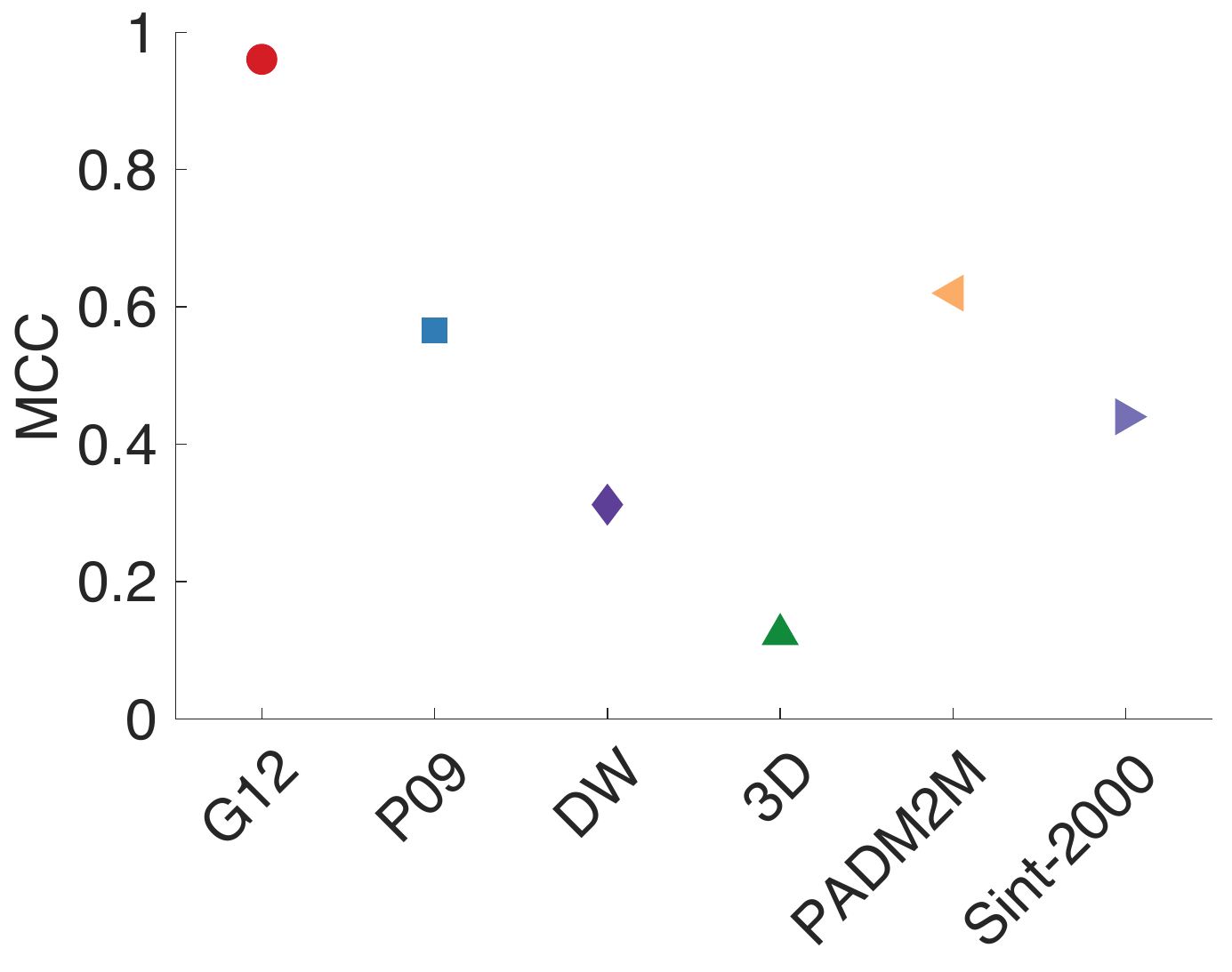}
	\caption{
	MCC of four models and two paleomagnetic reconstructions (PADM2M and Sint-2000).
	The optimal WT is computed using training data
	containing five events in the case of the models,
	and four events in the case of the paleomagnetic reconstructions
	(see text for details).
		}
	\label{fig:Skill_ModelsVsData}
\end{figure}

We also compute optimal MCC scores for PADM2M and Sint-2000 via training and verification.
We use the first 0.95 Myr, containing four events, for training (finding an optimal warning threshold)
and use the remaining 1.05 Myr, containing two events, for verification.
Table~\ref{tab:MainTable} lists these MCCs for PADM2M and Sint-2000,
together with the MCCs of the four models,
when computed with training data containing a comparable number of events 
(four events during training for the paleomagnetic data and five events
during training for the models, see Section~\ref{sec:RobustnessDurationOfTraining}).
Figure~\ref{fig:Skill_ModelsVsData} shows the (verification) MCC scores for PADM2M and Sint-2000 along with those of the models.
%The optimal WTs used here are obtained from training data that contain
%a comparable number of events (five events in the case of the models,
%see Section~\ref{sec:RobustnessDurationOfTraining}).

We first note from Table~\ref{tab:MainTable} that the MCC skill score drops from training to verification
and that the verification skills for PADM2M and Sint-2000 are quite different.
This is caused by the verification periods being extremely short,
with only two events during verification.
Thus, while the warning threshold we find from a limited observational record
may be quite accurate, it remains difficult to evaluate the skill of threshold-based predictions.
These difficulties are due to the shortness of the observational record -- 
we only have 2 Myr, with six events, to base our training \emph{and} validation on.
Nevertheless, we again find that paleomagnetic reconstructions tend to produce verification
MCC scores quite consistent  with what could be anticipated based on our analysis
of the ratio of the average decay time to the average event duration. The MCC associated with the paleomagnetic reconstructions,
indicative of the skill of intensity threshold based prediction, is indeed larger than
the MCC recovered for the 3D (smallest) and DW models, comparable to that of the P09 model,
and much smaller than that of the G12 model.  

\begin{figure}%[tb]
	\centering
	\includegraphics[width=.8\textwidth]{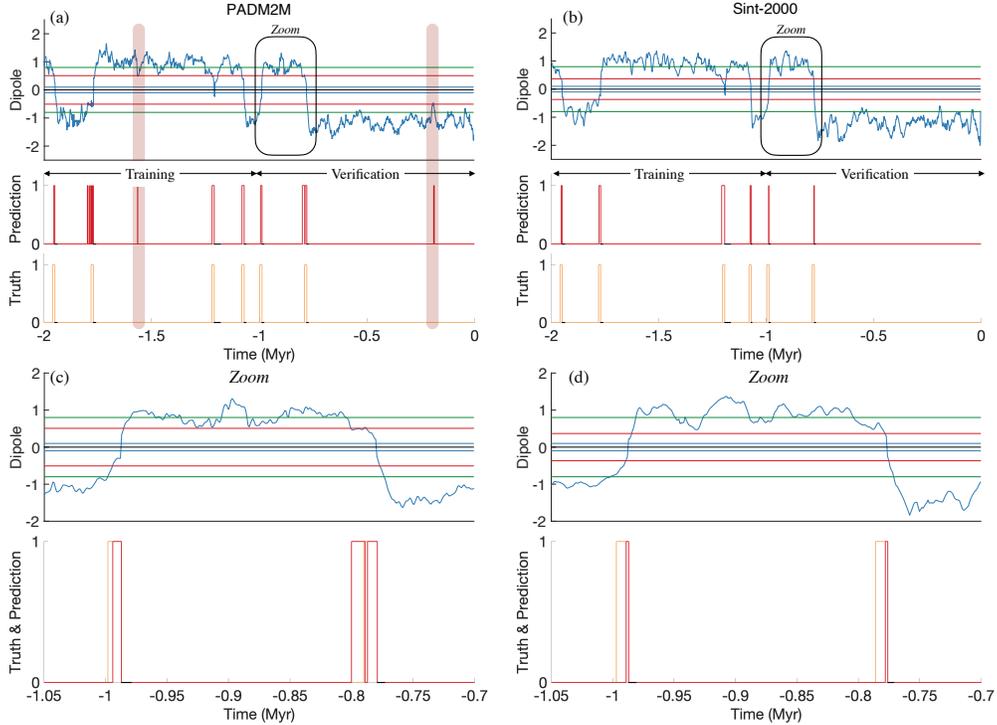}
	\caption{
	Illustration of threshold-based predictions for the PADM2M ((a) and (c)) 
	and Sint-2000 ((b) and (d)) reconstructions.
	The prediction horizon is $\text{PH} = 1$ (about 11 kyr)
	and the optimal warning threshold computed over 0.95 Myr of training data,
	containing four events.
	The corresponding warning thresholds (expressed in percent of the average intensity) are
	$\hat{\text{WT}}_\text{PADM2M} = 50.75\%$ ($2.70\cdot 10^{22}$ Am$^2$)
	and 
	$\hat{\text{WT}}_\text{Sint-2000}=36.75\%$ ($2.14\cdot 10^{22}$ Am$^2$)
	for respectively PADM2M and Sint-2000.
	Panels (a) and (b) contain three subfigures.
	\emph{Top}.
	Blue: dipole time series.
	Blue\slash green\slash red horizontal lines: 
	start-of-event\slash end-of-event\slash warning thresholds.
	\emph{Center}.
	Graphs are zero if the threshold-based prediction
	is ``no low-dipole event will start during the prediction horizon;''
	graphs are one if the threshold-based prediction
	is ``a low-dipole event will start during the prediction horizon.''
	\emph{Bottom}.
	Graphs are zero if no low-dipole event starts
	within the prediction horizon;
	graphs are one if a low-dipole event starts during the prediction horizon.
	Panels (c) and (d) show magnifications during a time interval that 
	includes the two reversals that occur during verification.
	}
	\label{fig:PredictionsData}
\end{figure}

We illustrate threshold-based predictions for the paleomagnetic reconstructions 
and $\text{PH} = 1$ in Figure~\ref{fig:PredictionsData}.
This figure first confirms that the choice of the start-of-event (ST) and end-of-event (ET) thresholds properly identifies
the six events of interest. 
There are five reversals and one major excursion, which corresponds to what is known
as the Cobb mountain subchron at 1.19 Myr, and is indeed an event during
which the field temporarily changed its polarity at low intensity.
This figure also illustrates the limitations of threshold-based predictions when using
PADM2M or Sint-2000. 
In the case of PADM2M, 
with an optimal warning threshold of $\hat{\text{WT}}_\text{PADM2M} = 50.75\%$
(corresponding to $2.70\cdot 10^{22}$ Am$^2$),
we note the occurrence of two instances of false positives,
where no low-dipole event is observed, but a low-dipole event is predicted,
(near the -1.5 Myr and -0.25 Myr marks).
One of these instances of false positives occurs during training, the other during verification.
Such false positives do not occur in the case of the Sint-2000,
which also has a lower optimal WT of $\hat{\text{WT}}_\text{Sint-2000}=36.75\%$
(corresponding to $2.14\cdot 10^{22}$ Am$^2$).
One may thus intuitively expect that the predictions will
have a lower skill when applied to PADM2M than to Sint-2000,
but in fact this is not the case:
the skill during verifications is higher for PADM2M than for Sint-2000,
but the skill for training is higher for Sint-2000 than for PADM2M. 
This is perhaps counter intuitive 
because one is tempted to think of false positives that occur ``far'' from a reversal
as more severe than false positives or false negatives that occur ``close'' to a reversal.
The MCC score, however,
does not apply special meaning to the categories of ``positive'' and ``negative,''
so that predictions of the timing of the two reversals, e.g., during verification, are 
more accurate for PADM2M than for Sint-2000.

The ROC curves and the MCC skill scores for the paleomagnetic reconstructions and models suggest
that the predictive skill of threshold-based predictions of the paleomagnetic reconstructions 
may be comparable to the skill of these predictions for the P09 model.
Because it is difficult to verify threshold-based predictions using the observational record only,
we may use the P09 model to investigate the skill of threshold-based predictions,
applied to the paleomagnetic reconstructions.
Threshold-based predictions ($\text{PH}=1$) for the P09 model are illustrated in 
Figure~\ref{fig:Prediction_Illu} (note that the predictions in Figure~\ref{fig:Prediction_Illu}
make use of a large training data set). 
Indeed, 
when training threshold-based predictions for P09 with training data 
that contains five low-dipole events (comparable to paleomagnetic reconstructions),
the optimal WT of the P09 model of $54.5\%$ is quite comparable to that
obtained for PADM2M ($50.75\%$) and slightly more than that obtained for Sint-2000 ($36.75\%$),
all of which are consistent with the range of values found in Figures~\ref{fig:Robustness_Training}
and~\ref{fig:Robustness_Training_MCC}.   
We also observe that the predictions for P09 are similar to the predictions for the paleomagnetic reconstructions.
We encounter a large number of true negatives,
several false negatives, 
for which the threshold-based predictions trigger a little too late,
and occasionally encounter false positives that 
occur during periods when no low-dipole event occurs.

In summary, 
we conclude that threshold-based predictions are feasible
for the paleomagnetic reconstructions,
but lead to moderate success.
They share similar characteristics as threshold-based predictions for the P09 model,
and suffer from similar caveats:
\begin{enumerate}[(i)]
\item
Low-dipole events can be predicted only a relatively short time ahead,
i.e., the prediction horizon should be about one average event duration or less.
On Earth's time scale, this means the prediction horizon should be about 10 kyr or less.
\item
Low-dipole events may be predicted a few kyr too late (false negatives),
which is significant in view of the relatively short prediction horizon.
\item
One must be prepared for false positives to occur 
even when no low-dipole event is about to happen.
\end{enumerate}

The above conclusions are supported by two paleomagnetic reconstructions, PADM2M and Sint-2000,
but threshold-based predictions show some sensitivity to which reconstruction we use.
This is perhaps best illustrated by the predictions in Figure~\ref{fig:PredictionsData},
but it is also clear from the skill scores in Table~\ref{tab:MainTable}.
As indicated above, differences between results stemming from PADM2M or Sint-2000
establish an uncertainty that cannot be resolved, 
because this uncertainty is caused by our limited ability to observe 
Earth's dipole over millions of years.
In this context,
we wish to point out that we did not use other global models,
e.g., PISO-1500 \citep{PISO1500},
because it is biased towards the North Atlantic region
due to the fact that only stacks with a high sedimentation rate are used 
(see, e.g., Figure 5 of \cite{PKC19}).
Indeed, PISO-1500 is less representative of the (global) axial dipole field than
PADM2M and Sint-2000 (see also \cite{PADM2M}).
Exploring the consequences of such differences is beyond the scope of our work.

Finally, we want to bring a few details to the reader's attention.
In particular, we want to emphasize that the average dipole intensity 
and the average event duration
for threshold-based predictions for PADM2M or Sint-2000
are computed using the entire 2 Myr record.
One could also envision to compute the average intensity 
based on training data only.
We decided not to do so for the following reasons.
The average intensity defines the start and end of an event,
because start-of-event 
and end-of-event thresholds are defined in terms of the average intensity.
The average event duration, and even the number of events,
are implicitly defined by the start-of-event
and end-of-event thresholds and, thus, also depends on the average intensity.
The average event duration, in turn,
is used in the definition of the prediction horizon.
In summary, the average intensity directly affects
(i) the number of events; (ii) the average event duration;
and (iii) the prediction horizon.
By computing the average intensity over the 2 Myr reconstructions,
we have assumed these difficulties away
and fix the average intensity a priori.
We find that this is more practically relevant
because the average intensity may be determined
by using additional information.
Nonetheless, we also made threshold-based predictions
for which we compute the average event duration based on training data
and the results are nearly identical to the results we show above.

\section{Concluding comments}
The main purpose of this study is to test the possibility that a low value of the axial dipole
intensity could be used as a natural indicator of an upcoming dipole reversal.
To answer this question, we analyzed a hierarchy of numerical models,
and Earth's axial dipole field as documented by the PADM2M and Sint-2000  
paleomagnetic VADM reconstructions \citep{PADM2M,SINT2000}.
More specifically, we test the possibility of relying on an intensity threshold-based strategy,
whereby once the axial dipole intensity 
drops below a warning threshold, it is predicted that the intensity will drop further
and lead to a low-dipole event (either a reversal or a major excursion) within some
specified time, called the prediction horizon.
Although the principle of such a strategy appears to be fairly intuitive,
implementing it in a robust way led us to introduce a dedicated methodology.

Our method requires that we define a warning threshold (WT), 
a start-of-event threshold (ST), an end-of-event threshold (ET)
and a prediction horizon (PH).
Both ST and ET appear to be most conveniently defined in terms 
the average intensity of the axial dipole 
(in practice ST=10\% and ET=80\%).
ST and ET also define an average event duration
(AED, average time elapsed between when the intensity passes below the ST and when it recovers back to above the ET).
%average time elapsed between the time the intensity reached the ST 
%and the time it reaches back the ET).
The prediction horizon is defined in terms of the average event duration 
and we consider predictions with PHs of about one AED.
Having chosen the ST, ET and PH,
we identify the warning threshold (WT) by maximizing a skill score.
Several skill scores have been tested, 
and all adequate choices led to similar conclusions.
Similarly, we showed that the exact choices of the ST and ET percentages are not critical,
provided these properly bracket the events of interest.
The code we use to implement the prediction
is available on github (\href{https://github.com/kjg136/Threshold}{https://github.com/kjg136/Threshold}).
We archived the code used to generate all figures
in (\href{https://doi.org/10.5281/zenodo.4267116}{https://doi.org/10.5281/zenodo.4267116}).

A first major conclusion is that the skills of intensity threshold-based predictions vary surprisingly widely
within the hierarchy of numerical models we investigated (G12, P09, DW and 3D models).
The only model that leads to a high skill (implying that the intensity threshold-based
predictions are reliable) is the G12 model.
This result is in line with the results obtained by \citet{MFH17},
who identified a high skill of intensity threshold-based predictions for this model,
using a simpler strategy and a less robust analysis.
All other models lead to lower skills,
implying that the intensity threshold-based predictions are less reliable.
This is, again, consistent with \citet{MFH17}, who investigated the P09 model
and a model (B13, \citet{Buffett2013}) similar to the DW model, but did not investigate the 3D model.
In the present study, we were able to rank these skills more accurately and identify one key property
that may play a major role in defining the skills of threshold-based predictions
in the context of numerical dynamos and VADM reconstructions (PADM2M and Sint-2000).

This key property is that skills of intensity threshold-based predictions
correlate with the ratio of the average decay time (defined as the time between the start of the event
and the most recent time instance at which the intensity is equal to the end-of-event threshold) 
to the average event duration.
The larger this ratio, the better the skill.
The models and the PADMD2M and Sint-2000 reconstructions are consistent with this rule.
%Specifically, the skills of the threshold-based predictions applied to paleomagnetic data are different from 
%the skills associated with the G12 model (higher skill),
%the DW model (lower skill) and the 3D model (surprisingly poor skill).
%The skills of predictions applied to paleomagnetic data are comparable
%to the skills associated with the P09 model.
%Similarly, the ratios of the average decay time to the average event duration for the paleomagnetic data 
%(about two to three)
%also are only comparable to the ratio associated with P09.
As already noted, this asymmetry between the way the field decreases towards a reversal and the way it recovers its strength
after the reversal is a well-known property of the field \citep{SINT2000},
which in fact may be related to the more general tendency of the Earth's magnetic field to spend more time decreasing
than increasing at any time \citep[see, e.g.,][]{ziegler2011asymmetry, avery2017asymmetry}.
What the present study thus suggests is that this slight asymmetry is what defines the 
skill of intensity threshold-based predictions when applied to Earth's magnetic field.
Unfortunately, because this ratio is about two to three, the skill
of threshold-based predictions is limited.
As our study further shows, this, more than the relatively short duration
of the Sint-2000 and PADMD2M reconstructions, 
is what likely makes intensity threshold-based predictions using these data modestly reliable.

Despite the limitations we identified for intensity threshold-based predictions, it is worth pointing out that
today's axial dipole field, with a magnitude of about $7.8\cdot 10^{22}$Am$^2$ \citep{Constable:2006},
is  much larger than the warning thresholds we identified by using either Sint-2000
($\hat{\text{WT}}_\text{Sint-2000}=36.75\%$ 
of the average $5.81\cdot 10^{22}$ Am$^2$, 
which amounts to $2.14\cdot 10^{22}$ Am$^2$)
or PADM2M 
($\hat{\text{WT}}_\text{PADM2M}=50.75\%$ 
of the average $5.32\cdot 10^{22}$ Am$^2$, 
amounting to $2.70\cdot 10^{22}$ Am$^2$).
Intensity threshold-based predictions thus suggest that 
no low-dipole event will occur within the next 10 kyr.
This is in line with many other recent predictions,
see, e.g., \cite{Constable:2006, MFH17, BKHWG18}.

As an interesting additional outcome of this study, we note that testing the skills of 
threshold-based predictions on numerical dynamos 
is a fairly discriminating way of testing the Earth-like
nature of the axial dipole field behavior of the models. 
This skill is distinct from the ability of numerical
simulations to reproduce the frequency with which reversal occurs.
This is evident from the fact that threshold-based predictions have different skills for the DW and P09 models,
whereas both models are characterized by reversal frequencies comparable to that of the Earth
over the last 25 My (about 5 reversals per Myr). 
As this skill appears to be correlated
with the ratio of the average decay time to the average event duration (a measure of the asymmetry
with which the field evolves towards a reversal and next recovers its full strength),
it also appears to be distinct from other criteria often used to characterize the Earth's dipole field behavior,
such as its frequency content  \citep{CJ05}, or the relative time spent in transitional periods
(based on dipole latitudes being less than $45^\circ$), as recently suggested by \citet{sprain2019assessment}.
%%%%
 Furthermore, in spite of its favorable ratings according to the criteria 
 defined by \citet{Christensen2010} for the recent field, 
 and \citet{sprain2019assessment} for the paleomagnetic field (recall section~\ref{sec:3d_pres}), the
 field produced by the 3D model appears to not match
that of the Earth's field (as described by PADM2M and Sint-2000)
in terms of intensity threshold-based prediction
skill (and ratio of the average decay time to the average event duration).
In agreement with the suggestions of \citet{ziegler2011asymmetry} and \citet{avery2017asymmetry}, 
and since it appears to play a significant role in the way reversals occur, 
we strongly encourage the community to also consider predictive skills 
and asymmetric temporal behavior
as additional criteria to identify Earth-like dynamo simulations.

The present study shows that intensity threshold-based predictions of reversals appear to be of only limited value, but we emphasize that we investigated 
these limitations for only one specific threshold-based prediction, 
namely predicting whether a reversal or major excursion occurs during a specified time window.
Other types of predictions might deal instead with the probability of a reversal or major excursion during a specified time window. 
In this case, a large number of reversals would be needed to test these predictions.
It is also worthwhile to comment on other routes to more robust and reliable predictions.
Taking advantage of machine-learning and deep learning could be a possibility \citep{GBC2016}.
In this context, however, one should be careful to check that the shortness of the paleomagnetic reconstructions
is not a limiting factor, as deep learning is known 
to work best when data availability is vast,
and only poorly when data are limited. 
Another approach is to rely on merging the observations
in a process called data assimilation (DA) \citep[see, e.g.,][]{CBBE18}.
This strategy has been successful in numerical weather prediction
because the atmospheric model is of high quality, 
and because observations of the atmospheric state are plentiful \citep{BTG15}. It currently is developing 
in the field of geomagnetism \citep{Fournier2010}. 
Using DA for predicting dipole reversals, however, is difficult due to 
the lack of a suitable 3D model that can be run fast enough 
and the fact that the observations are limited to 
the virtual axial dipole moment over 2 Myr.
Here, the main difficulty lies in identifying, or creating, useful models
that are simple enough to allow for data assimilation
but complex enough to represent all relevant time scales.
Nevertheless, \citet{MFH17} recently showed that using such an approach with the G12 model 
and assimilating either PADM2M or Sint-2000, could lead to some success.
No similar success could be reached with the P09 model, which was also tested.
In that approach, indeed, the key to success appears to be the dynamical way the axial dipole
produced by the model approaches reversals.
It appears that the way the G12 model approaches reversals is more 
similar to how Earth's axial dipole field approaches reversals,
than the P09 model.
This leads to the interesting possibility of finding a better suited low-dimensional model with properties intermediate between the G12 model (whose decay-time properties make it well suited for data assimilation) and P09 (with intensity threshold-based prediction properties closest to that of the paleomagnetic reconstructions) leading to better predictions of reversals several kyr ahead.
%This leads to the interesting conclusion that an even better suited low-dimensional model
%with properties intermediate between model G12 (best for data assimilation) and P09 (with intensity threshold-based prediction properties closest to that of the paleomagnetic data) could possibly be found to even better predict reversals several kyr ahead.

\section*{Acknowledgements}
KG acknowledges that this work was supported by NASA Headquarters under the NASA Earth and Space Science Fellowship Program - Grant ``80NSSC18K1351''.
This work was supported in part by the French Agence Nationale de la Recherche under grant ANR-19-CE31-0019 (revEarth).
All authors would like to thank Nathanael Schaeffer
(ISTerre, CNRS, Universit\'e Grenoble Alpes) 
and Thomas Gastine 
(Universit\'e de Paris, Institut de Physique du Globe de Paris)
for allowing us to use the dipole time series of the 3D model.
We acknowledge GENCI for access to the Irene resource (TGCC) under
grants ``Grand Challenge'' GCH0315 and A0060407382. 
We thank Maggie S. Arvery (UC Berkeley) and an anonymous reviewer for 
helping us improve this paper.
We thank Cathy Constable (Scripps Institution of Oceanography,
University of California, San Diego) 
and Bruce Buffett (UC Berkeley) for meaningful discussions.
AF thanks Richard Bono and Courtney Sprain for their  assistance in the calculation of $\Delta Q_{PM}$.
All authors contributed to the ideas behind the approach taken in the manuscript
and all authors contributed to writing the paper; 
KG wrote the code.

%\emph{Code availability}: 
%The code is available at TBD.%, which has DOI:10.5281/zenodo.1442345.

\bibliographystyle{./gji-latex/gji}
\bibliography{Refs/references}

%\newpage
%\bibliographystyle{./gji-latex/gji}
%\bibliography{Refs/references}

%Tables
\newpage
%\begin{table}%[tb]
%\label{tab:Skewness}
%\caption{Skewness of the models' intensity distributions.}
%\begin{center}
%\begin{tabular}{cccc}
%G12 & P09 & DW  & 3D \\\hline
%0.92 &-0.94 &-0.17 &-0.46
%\end{tabular}
%\end{center}
%\end{table}%

%Figures

\newpage

\appendix
%\section{Acronyms}
\begin{table}
\caption{Acronyms used in this paper}
\begin{center}
\begin{tabular}{lll}
Type & Acronym & Explanation\\\hline\hline
\multirow{2}{*}{Outcomes of events} 
&P& Number of positives\\
&N & Number of negatives\\\hline

\multirow{4}{*}{Outcomes of predictions} 
&TP & True positive\\
&FP& False positive \\
&TN & True negative \\
&FN & False negative \\\hline

\multirow{3}{*}{Receiver operator charecteristics} 
& TPR & True positive rate (equation~(\ref{eq:TPRFPR}))\\
& FPR & False positive rate (equation~(\ref{eq:TPRFPR}))\\
& ROC & Receiver operator characteristic\\\hline

\multirow{4}{*}{Skill scores} &
 ACC & Accuracy  (equation~(\ref{eq:ACC}))\\
& $\F$ & $\F$ skill score (equation~(\ref{eq:F1}))\\
& CSI & Critical success index (equation~(\ref{eq:CSI}))\\
& MCC & Mathews correlation coefficient (equation~(\ref{eq:MCC}))\\\hline

\multirow{4}{*}{Threshold-based predictions} &
   ST & Start-of-event threshold\\
& ET & End-of-event threshold\\
& WT & Warning threshold\\
& PH & Prediction horizon\\
& AED & Average event duration \\
& ADT & Average decay time \\
&$\rho=\frac{\text{ADT}}{\text{AED}}$& ratio of ADT and AED\\\hline

\multirow{7}{*}{Models} 
&G12 & Differential equation model \citep{G12}\\
&P09 & Stochastic model  \citep{PFDV09}\\
& DW & Stochastic double well model \citep{MB19}\\
& 3D model & 3-dimensional dynamo simulation (unpublished)\\
& SDE & Stochastic differential equation\\
& MHD & Magneto-hydrodynamic \\
& DA & Data assimilation \\\hline

\multirow{3}{*}{Data} &
 VADM & Virtual axial dipole moment\\
&PADM2M &  VADM reconstruction \citep{PADM2M}\\
&Sint-2000 &  VADM reconstruction \citep{SINT2000}
\end{tabular}
\end{center}
\label{tab:AccronymTable}
\end{table}%

\end{document}